\documentclass[useAMS,usenatbib]{mn2e}

\input{epsf}
\usepackage{graphicx}
\usepackage{rotating}

\usepackage{lscape}
\usepackage{multirow} 

\newcommand{\mic}{\,$\mu$m}
\newcommand{\kms}{\,km s$^{-1}$}
\newcommand{\dg}{$^{\circ}$}

\newcommand{\Msol}{\,M$_{\odot}$}

\newcommand{\htwo}{H$_2$ 2.122\,$\mu$m}
\newcommand{\hco}{H$^{13}$CO$^+$}

\title[GBS HARP Survey: Taurus]
{The JCMT Legacy Survey of the Gould Belt: a first look at Taurus with HARP}

\author[C.J.~Davis et al.]
       {C.J.~Davis$^{1}$, 
        A.~Chrysostomou$^{1}$,
        J.~Hatchell$^{2}$, 
        J.G.A.~Wouterloot$^{1}$, 
        J.V.~Buckle$^{3,16}$, \newauthor
        D.~Nutter$^{4}$,
        M.~Fich$^{17}$,
        C.~Brunt$^{2}$,
        H.~Butner$^{5}$, 
        B.~Cavanagh$^{1}$,
        E.I.~Curtis$^{3,16}$,   \newauthor   
        A.~Duarte-Cabral$^{6}$,                  % Observer - 2008-01-12
        J.~Di Francesco$^{7,9}$,                   % Observer - 2007-11-23/24/25
        M. Etxaluze$^{8,15}$, 
        P.~Friberg$^{1}$,                        % Observer - 2008-09-19
        R.~Friesen$^{7,9}$,\newauthor
        G.A.~Fuller$^{6}$, 
        S.~Graves$^{3,16}$,
        J.S.~Greaves$^{10}$,
        M.R.~Hogerheijde$^{11}$,
        D.~Johnstone$^{7,9}$,\newauthor
        B.~Matthews$^{7}$, 
        H.~Matthews$^{7,12}$,                     % Observer - 2007-11-26        
        J.M.C.~Rawlings$^{13}$,
        J.S.~Richer$^{3,16}$,
        J.~Roberts$^{14}$,  \newauthor
        S.~Sadavoy$^{7,9}$,                    % Observer - 2007-11-23/24/25
        R.J.~Simpson$^{4}$,                     % Observer - 2007-11-23/24/25
        N.~Tothill$^{2}$,                       % Observer - 2008-08-20/21/23/24
        Y.~Tsamis$^{18}$,
        S.~Viti$^{13}$, \newauthor  
        D.~Ward-Thompson$^{4}$, 
        Glenn J.~White$^{8,15}$ \&
        J.~Yates$^{13}$
           \\ \\
$^1$Joint Astronomy Centre, 660 North A'oh\={o}k\={u} Place, University Park, Hilo, 
      Hawaii 96720, U.S.A. \\
$^2$School of Physics, University of Exeter, Stocker Road, Exeter, U.K. \\
$^3$Cavendish Laboratory, Cambridge University, J.J. Thomson Avenue, Cambridge, CB3 9HE, U.K. \\
$^4$School of Physics and Astronomy, Cardiff University, 5 The Parade, Cardiff, U.K. \\
$^5$Dept. of Physics and Asrtronomy, James Madison University, 901 Carrier Dr., Harrisonburg, 
       VA 22807, U.S.A. \\
$^6$Jodrell Bank Centre for Astrophysics, School of Physics and Astronomy, Alan Turing
       Building, The University of Manchester, \\ Manchester, M13 9PL, U.K. \\
$^7$Herzberg Institute of Astrophysics, National Research Council of Canada, 5071 W. Saanich Rd., 
       Victoria, BC, Canada.\\
$^8$Department of Physics and Astronomy, Open University, Walton Hall, Milton Keynes, U.K. \\
$^9$Department of Physics and Astronomy, University of Victoria, 3800 Finnerty Road, Victoria, BC,
       Canada \\
$^{10}$Scottish Universities Physics Alliance, Physics and Astronomy, University of St. Andrews, 
       North Haugh, St. Andrews, U.K. \\
$^{11}$Leiden Observatory, Leiden University, PO Box 9513, 2300 RA, Leiden, the Netherlands \\
$^{12}$Dominion Radio Astrophysical Observatory, National Research Council of Canada, 
       White Lake Road, Penticton, Canada. \\
$^{13}$Department of Physics and Astronomy, University College London, Gower Street, London, U.K. \\
$^{14}$Centro de Astrobiologia (CSIC/INTA), Instituto Nacional de Tecnica Aeroespacial, 
        Ctra. de Torrejon a Ajalvir km 4, E-28850 \\ Torrejon de Ardoz, Madrid, Spain. \\
$^{15}$Science and Technology Facilities Council, Rutherford Appleton Laboratory, Chilton, 
        Didcot, U.K. \\
$^{16}$Kavli Institute for Cosmology, c/o Institute of Astronomy, University of Cambridge,
Madingley Road, Cambridge, CB3 0HA, U.K. \\
$^{17}$Department of Physics and Astronomy, University of Waterloo, Waterloo, Ontario, Canada. \\ 
$^{18}$Instituto de Astrofísica de Andaluc\'ia (CSIC), Camino Bajo de Hu\'etor, 50, 18008 
Granada, Spain.
        }

\begin{document}

%\date{Accepted 2009 ... ;  Received 2009 ... ; in original form 2009 ... }
%\pagerange{\pageref{firstpage}--\pageref{lastpage}} \pubyear{2008}

\maketitle

\label{firstpage}

\begin{abstract} 
As part of a JCMT Legacy Survey of star formation in
the Gould Belt, we present early science results for Taurus.  
CO J=3-2 maps have been secured along the north-west ridge
and bowl, collectively known as L~1495, along with deep $^{13}$CO and
C$^{18}$O J=3-2 maps in two sub-regions.  With these data we search
for molecular outflows, and use the distribution of flows, HH objects
and shocked H$_2$ line emission features, together with the
population of young stars, protostellar cores and
starless condensations to map star formation across this extensive
region.  In total 21 outflows are identified.
%% most are associated with dense
%% molecular cores and embedded (Class 0/I) protostars rather than (Class
%% II) T Tauri stars; H$_2$ line emission features are also excited in
%% some flows.  
It is clear that the bowl is more evolved than the
ridge, harbouring a greater population of T Tauri stars and a more
diffuse, more turbulent ambient medium.  By comparison, the ridge
contains a much younger, less widely distributed population of
protostars which, in turn, is associated with a greater number of
molecular outflows.  We estimate the ratio of the numbers of
prestellar to protostellar cores in L~1495 to be
$\sim$1.3--2.3, and of gravitationally unbound starless
cores to (gravitationally bound) prestellar cores to be
$\sim$1.  If we take previous estimates of the protostellar lifetime
of $\sim$5$\times$10$^5$~yrs, this indicates a prestellar
lifetime of 9($\pm$3)$\times$10$^5$~yrs.  
%% Since most of the starless
%% cores are likely to be prestellar, the prestellar collapse time must
%% be at least as long as the Class 0/I protostellar lifetime. 
From the number of outflows we also crudely estimate the star formation
efficiency in L~1495, finding it to be compatible with a canonical
value of 10-15\%.  We note that molecular outflow-driving sources have
redder near-IR colours than their HH jet-driving counterparts. We also
find that the smaller, denser cores are associated with the more
massive outflows, as one might expect if mass build-up in the flow
increases with the collapse and contraction of the protostellar
envelope.
\end{abstract}


%%%%%%%%%%%%%%%%%%%%%%%%%%%%%%%%%%%%%%%%%%%%%%%%%%%%%%%%%%%%%
%%%%%%%%%%%%%%%%%%%%%%%%%%%%%%%%%%%%%%%%%%%%%%%%%%%%%%%%%%%%%

\section{Introduction}

The Gould Belt Survey \citep[GBS:][]{war07a} is a large, legacy
programme currently underway at the James Clerk Maxwell Telescope
(JCMT).  The goal is to gather both heterodyne line observations and
submillimetre (submm) continuum images of nearby star forming regions.
The survey targets the Gould Belt, which is a ring of O-type
stars and molecular clouds with a radius of $\sim$350~pc. It is
inclined at $\sim$20\dg\ to the Galactic Plane and is centred on a
point 200~pc from the Sun \citep[ascending node, $\Omega =
275-295$;][]{tor00}.  The Gould Belt includes well known regions such
as Orion, Taurus-Auriga-Perseus, Serpens, Lupus, Ophiuchus and the
Pipe nebula.  Taurus is one of four regions initially targeted in
$^{12}$CO J=3-2 (hereafter CO 3-2), $^{13}$CO 3-2, and C$^{18}$O
3-2. These four regions -- Orion A \citep{buc10}, Serpens
\citep{gra10}, Ophiuchus (White et al., in prep.) and Taurus (this
paper) -- were selected as being complementary, active, nearby,
relatively well known, and well studied. Extensive submm
continuum mapping, at 450\mic\ and 850\mic , is also planned as part
of the survey
\citep{war07a}.

In studies of nearby, low-mass star forming regions CO 3-2 emission
samples the denser, warmer gas which often envelopes cores where new
stars are forming (temperatures $\sim 10-50$~K; gas densities of the
order of $10^4 - 10^5$~cm$^{-3}$).  CO 3-2 is also a proven tracer of
outflow activity (e.g. Hatchell, Fuller \& Ladd 1999; Knee \& Sandell
2000; Davis et al. 2000a; Hatchell, Fuller \& Richer 2007a; Lee et
al. 2007; Bussmann et al. 2007; Yeh et al. 2008). With the JCMT,
moderate spatial resolution is attainable at 345~GHz ($\sim$14\arcsec
), which often allows one to disentangle multiple outflows in
clustered regions.  The higher critical density for excitation of the
3-2 line (over, say, the 1-0 emission line) can also lead to better
maps of the dense, more collimated flow components. Molecular outflows
are known to emanate from the very youngest sources, objects that are
still accreting most of their final mass.  New, large-scale surveys in
CO can therefore be used to identify the locations of these embedded
protostars, leading to a more complete census of the youngest stars,
and an indication of the overall youth and star formation activity in
a region \citep[e.g.][]{hat07a, dav08, hat09, buc10}.

%%%%%%%%%%%%%%%%%%%%%%%%%%%
%%% Figure  %%%%%%%%%%%%%%%
\begin{figure*}
\centering
\epsfxsize=17cm
%\epsfbox{/export/data/cdavis/jcmt/TAURUS/PAPER-FIGURES/overview2.eps}
\epsfbox{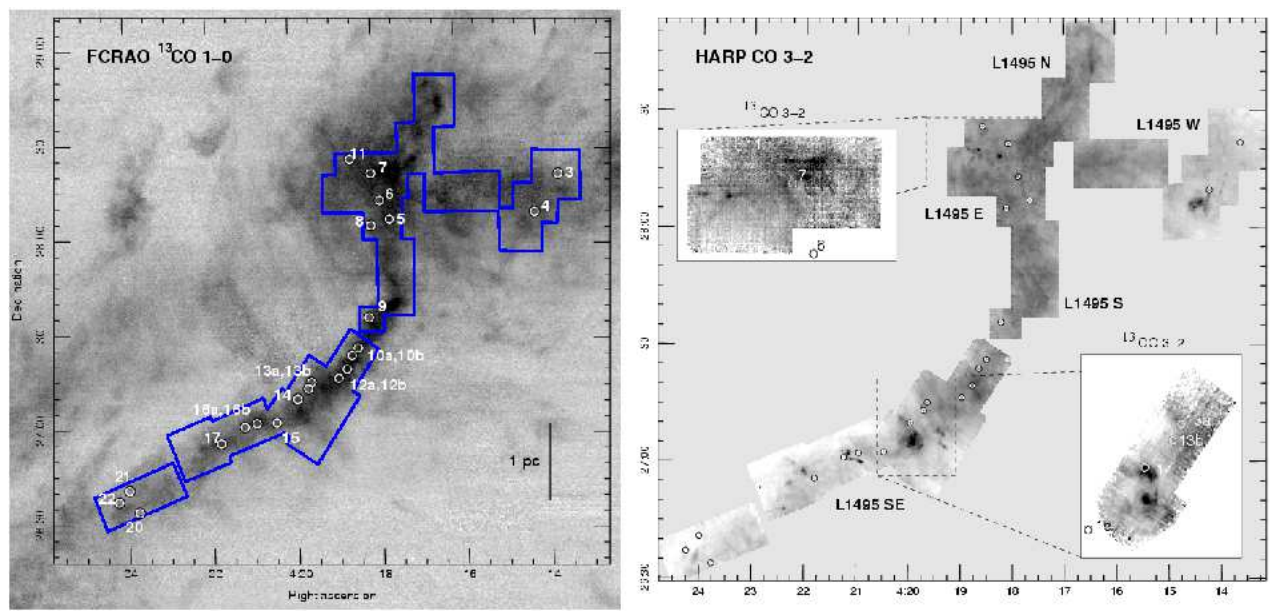}
\caption[] {{\em Left -} an overview of L~1495 in the north-west corner of Taurus. 
The image shows integrated $^{13}$CO 1-0 emission from the FCRAO 
survey of \citet{gol08}.  The positions of the \hco\ cores identified by 
\citet{oni02} are marked with circles and labelled using their 
core numbering scheme. {\em Right - } a complete map
of all regions observed with HARP in CO 3-2 (intensity integrated between
2.0 and 11.0~\kms), with the positions of the \hco\ cores again
marked. The areas mapped in $^{13}$CO and C$^{18}$O 3-2 are also
indicated; maps of $^{13}$CO emission, integrated between 4.9 and
9.1~\kms (L~1495~E) and 4.5 and 8.2~\kms (L~1495~SE), are shown inset
as grey-scale images. In the main text we refer to L~1495 N, E, S and W as the bowl, 
and L~1495~SE as the south-east ridge. }
\label{over}
\end{figure*}
%%%%%%%%%%%%%%%%%%%%%%%%%%%

%%%%%%%%%%%%%%%%%%%%%%%%%%%
%%% Table   %%%%%%%%%%%%%%%
\begin{table*}
\centering
\begin{minipage}{195mm}
 \caption{Log of CO 3-2 observations in L~1495.}
 \begin{tabular}{@{}cclcccccr@{}} 

 \hline
Region$^a$ & Area$^a$ & Cube$^a$ & Date$^a$ & RA$^b$  & Dec$^b$ & PA$^c$ & Map size$^c$ & $v_o$$^d$ \\
           &          &          &          &(J2000.0)&(J2000.0)&        &              &   (\kms ) \\
 \hline
 
Bowl & L~1495~E   & L~1495     & 2007-11-24 & 4:18:04.8 & 28:21:45 &  90\dg & 20\arcmin$\times$10\arcmin & 6.6-7.1\\  
$''$ & $''$       & L~1495-S   & 2007-12-22 & 4:17:50.0 & 28:09:00 &   0\dg & 10\arcmin$\times$15\arcmin & 5.9-6.7\\  
$''$ & $''$       & L~1495-S2  & 2008-09-17 & 4:18:48.0 & 28:14:30 &   0\dg & 10\arcmin$\times$10\arcmin & $\sim$5.9-6.7\\ 

$''$ & L~1495~N   & L~1495-NE1 & 2008-11-08 & 4:16:59.0 & 28:30:13 &   0\dg & 10\arcmin$\times$14\arcmin & $\sim$6.7\\ 
$''$ & $''$       & L~1495-NE2 & 2009-01-30 & 4:16:28.1 & 28:45:33 &   0\dg & 10\arcmin$\times$14\arcmin & $\sim$6.7\\
 
$''$ & L~1495~S   & L~1495-S3  & 2008-11-08 & 4:17:34.0 & 27:49:20 &   0\dg & 10\arcmin$\times$22\arcmin & 6.6-6.8\\ 
$''$ & $''$       & L~1495-SE1 & 2008-03-29 & 4:18:09.1 & 27:35:17 &   0\dg &  5\arcmin$\times$5\arcmin  & 5.6\\ 

$''$ & L~1495~W   & L~1495-W   & 2008-08-21 & 4:14:10.0 & 28:12:00 &   0\dg & 10\arcmin$\times$10\arcmin & 6.7\\
$''$ & $''$       & L~1495-W1  & 2009-01-31 & 4:15:54.7 & 28:16:33 &   0\dg & 22\arcmin$\times$10\arcmin & $\sim$6.7\\ 
$''$ & $''$       & L~1495-W2  & 2008-10-07 & 4:14:34.0 & 28:04:00 &   0\dg & 13\arcmin$\times$13\arcmin & 6.7\\ 
$''$ & $''$       & L~1495-W3  & 2009-01-30 & 4:13:35.0 & 28:22:13 &   0\dg & 13\arcmin$\times$13\arcmin & 6.7\\ \\

Ridge& L~1495~SE  & L~1495-SE2 & 2008-03-15 & 4:18:41.3 & 27:22:00 & 147\dg &  8\arcmin$\times$15\arcmin & 5.3-5.7\\
$''$ & $''$       & L~1495-SE3 & 2007-11-25 & 4:19:42.4 & 27:13:24 & 147\dg &  8\arcmin$\times$15\arcmin & 6.5\\
$''$ & $''$       & L~1495-SE3b& 2008-01-12 & 4:20:10.0 & 27:05:00 & 147\dg & 10\arcmin$\times$10\arcmin & 6.4\\
$''$ & $''$       & L~1495-SE3c& 2008-09-17 & 4:19:22.0 & 27:04:06 & 147\dg &  9\arcmin$\times$23\arcmin & $\sim$6.4\\
$''$ & $''$       & L~1495-SE4 & 2008-08-21 & 4:21:14.1 & 27:01:11 & 113\dg & 10\arcmin$\times$20\arcmin & 6.6-6.7\\
$''$ & $''$       & L~1495-SE4b& 2008-10-07 & 4:22:22.0 & 26:55:00 & 113\dg & 10\arcmin$\times$10\arcmin & $\sim$6.6\\ 
$''$ & $''$       & L~1495-SE5 & 2008-11-08 & 4:23:44.5 & 26:39:00 & 113\dg & 9\arcmin$\times$ 25\arcmin & 6.6-6.7\\ 

 \hline
 \label{obs}
 \end{tabular}

%\smallskip 
$^a$The areas in the bowl and ridge regions of L~1495 are made up of multiple data cubes, 
    often observed on different nights. \\  
$^b$Coordinates of the cube centres. \\
$^c$Cube map size and position angle of the map long axis (measured E
of N); with ``basket-weave'' mapping the scan direction is along \\
and orthogonal to this axis. \\
$^d$The LSR velocities of the H$^{13}$CO$^+$ cores found in or near (if an approximate 
value is given) each region by OMK02, which we adopt as the local ambient gas
velocity. \\
\end{minipage}
\end{table*}

%%%%%%%%%%%%%%%%%%

Taurus is a much-studied region of low mass star formation (see
Kenyon, G\'omez \& Whitney 2009, for a detailed review).  The
Taurus molecular cloud covers an area in excess of 100 deg$^2$
\citep{ung87,nar08,gol08}. LDN~1495 (L~1495 hereafter) is one of three
main areas associated with a high density of young stars.  L~1495 lies
in the north-west corner of Taurus and coincides with Barnard dark
nebulae B~7, 10, 209, 211, 213 and 216. In molecular line and
extinction maps L~1495 comprises an extended, knotty filament -- here
referred to as the ``south-east ridge'' -- and, at its northern
extremity, a more diffuse cloud -- the ``bowl''. 
\citet{gol08} refer to the ridge as B~231 and label only the bowl
L~1495.  From their extensive CO observations they estimate masses of
1095~\Msol\ and 2626~\Msol\ for the ridge and bowl, respectively, and
assign areas of 13.7~pc$^2$ (2.3 sqr degrees) and 31.7~pc$^2$ (5.3
sqr. degrees) for each region (see their Table 4).  They also note
that, quite remarkably, the SE ridge is 75\arcmin\ (3~pc) long yet
only 4.5\arcmin\ (0.2~pc) wide. A similar collimated feature in the 
NGC~6334 star forming region has recently been analysed by
\citet{mat08}. 

In this paper we split the bowl into four areas, L~1495~N, S, E and W,
and label the ridge L~1495~SE.  L~1495~E harbours the classical T
Tauri star (TTS) CoKu Tau-1 and HH~156 \citep{eis98}; L~1495~W (B~209)
is associated with the TTS CW~Tau (Elias~1, Hubble~4) which drives
HH~220 and possibly also HH~826-828 \citep{gom93, mcg04}.
HH~390/391/392 and HH~157 are located along the L~1495~SE ridge
(Gomez, Whitney \& Kenyon 1997; Eisl\"offel \& Mundt 1998); HH~157 is
driven by the TTS Haro~6-5B (FS~Tau~B) and comprises a spectacular HH
jet and bow shock that extend over 30\arcsec --40\arcsec .

L~1495~W has been mapped in C$^{17}$O 1-0, C$^{18}$O 1-0, CS 2-1,
N$_2$H$^+$ 1-0 (at FCRAO) and in NH$_3$ (1,1) and (2,2) emission
(at Effelsberg) by \citet{taf02} -- the region they label ``L~1495"
roughly coincides with L~1495~W.  N$_2$H$^+$ 1-0 maps of the south-east
ridge are presented by \cite{tat04}. More extensive $^{13}$CO and
C$^{18}$O 1-0 maps of Taurus are discussed by \citet{miz95}, and
\citet{oni96}; CO and $^{13}$CO 1-0 maps, covering 98 square degrees
with 45\arcsec\ resolution, have recently been published by
\citet{nar08} and \citet{gol08}.  Finally, extensive yet high spatial
resolution \hco\ 1-0 observations of the molecular cores in Taurus,
including the L~1495 ridge and bowl, are discussed by
\citet[][hereafter referred to as OMK02]{oni02}. These
\hco\ observations, to which we refer throughout the paper, were
obtained at the 45~m Nobeyama telescope with a beam width of 20\arcsec .

\citet{ken95} present a list of $\sim$300 Young Stellar Objects (YSOs)
in Taurus, derived from multi-wavelength observations and complemented
recently with data from the {\em Spitzer Space Telescope}
\citep[][though note that the {\em Spitzer} observations do not cover
L~1495~W or the bottom of the south-east ridge]{har05, luh06}. Many of
the youngest sources have been observed in the 1.3~mm continuum survey
obtained by \citet{mot01}. In L~1495 the young stars are clustered
toward the eastern and western sides of the L~1495 bowl, and are
tightly bound along the narrow, south-east ridge. The protostellar
population as a whole is discussed in the review of \citet{ken09}, who
also present an updated list of YSOs in Taurus.  Kenyon et al. note in
their article that classical TTSs typically have near-infrared
(near-IR) colours $H-K_s < 1-1.5$ and mid-infrared (mid-IR) colours
[3.6]-[4.5]$\sim$0.5-1.0 and [5.8]-[8.0]$\sim$0.25-0.75; embedded
protostars (Class 0/I sources) are expected to be redder.

In this paper we use an up-to-date list of ``Taurus Young Stars'',
kindly provided by K. Luhmann, to search for candidate outflow sources
in L~1495.  This list is essentially the same as that presented by
\citet{ken09}.  We also assume that ``starless'' cores are low mass
($M < 10$~\Msol ) dense cores without compact luminous sources of any
mass \citep{fra07}. ``Prestellar'' cores are a subset of starless
cores, since they must also be gravitationally bound \citep{war07b}.
Establishing whether a core is bound or not is observationally very
difficult (although the virial theorem gives some indication).
We therefore refer to cores that are not obviously associated with a
young star as starless; cores that do appear to contain a young star
are referred to as ``protostellar''.

Traditionally, a distance of 140~pc has been used for Taurus
\citep{ung87, eli78}, on the basis of estimates from star counts
\citep{mcc39}, the reddening of field stars as a function of 
distance \citep{got69}, and photometric distances measured 
to the exciting stars of reflection nebulae
\citep{rac68}.  Analysis of the velocity field in wide-field
molecular line studies suggests that ambient gas velocities change by
only a few \kms\ across Taurus, and are particularly limited across
L~1495 \citep[][see also Sect. 3.1 below]{ung87, nar08, gol08}.  This
supports a more-or-less common distance to the molecular clouds and
star forming regions in Taurus.  Recent trigonometric parallax
measurements suggest that the eastern portion of Taurus may be
furthest from us, at a distance of $\sim 161$~pc, with the south
(around T Tau) at an intermediate distance of 147~pc and the west,
L~1495, marking the near side of the cloud at a distance of 130~pc
\citep{tor09}. However, these distances are based on parallax
measurements for only a handful of stars. We therefore adopt the canonical
distance of 140~pc for L~1495, consistent with previous studies, most
notably \citet{gol08} and OMK02.

Overall, our goal with this paper is to search for outflows in
L~1495, identify their driving sources, and compare the properties of
these outflows with those of the driving sources and associated
molecular cores.

%%%%%%%%%%%%%%%%%%%%%%%%%%%%%%%%%%%%%%%%%%%%%%%%%%%%%%%%%%%%%%%%
%%%%%%%%%%%%%%%%%%%%%%%%%%%%%%%%%%%%%%%%%%%%%%%%%%%%%%%%%%%%%%%%

\section{Observations and Data Reduction}

\subsection{Data Acquisition}

As part of the JCMT Gould Belt Legacy Survey, the 325--375~GHz
heterodyne receiver array HARP was used to map 18 near-contiguous
regions across the L~1495 bowl and along the south-east ridge in CO
3-2 emission (Fig.~\ref{over} and Table~\ref{obs}), as well as two
smaller sub-regions in $^{13}$CO 3-2 and C$^{18}$O 3-2 emission.  HARP
is a 16 receptor (4$\times$4 array) single side-band SIS
receiver system acting as a front-end to the ACSIS digital
auto-correlation spectrometer \citep{hov00,smi03,buc09}.

CO 3-2 data at a rest frequency of 345.79599~GHz were acquired
in dual sub-band mode.  Maps with modest and high spectral resolution
were acquired simultaneously, with  1024 ACSIS channels covering a  
bandwidth of 1.0~GHz
   at a spectral resolution of 977~kHz (0.85 kms$^{-1}$),
and with  4096 channels covering a bandwidth of 250~MHz
   at a spectral resolution of 61~kHz (0.05 kms$^{-1}$).
Only the latter are presented in this paper. In separate observations,
$^{13}$CO 3-2 (330.58796~GHz) and C$^{18}$O 3-2 (329.33055~GHz) data
were acquired simultaneously in dual sub-band
mode using only the high spectral resolution set-up.  At these
frequencies the telescope Half Power Beam Width measures 14\arcsec ,
corresponding to 0.0095~pc at a distance of 140~pc.

Maps were obtained by scanning the telescope along rows parallel to
the sides of the pre-defined mapping areas, each row separated by half
or a quarter of an array to improve the spatial sampling. This map is
then repeated but by scanning in a perpendicular direction. This
strategy of ``basket weaving" helps to generate flatter maps and evens
out the noise in the final data cube. For most of the data presented
here, 2--4 receptors were inoperable; for those fields where 4
receptors were unavailable (regions S2, W2, SE3c and SE4b) the array
was stepped by a quarter of the array to prevent empty spaces or
poorly sampled regions in the final maps.

Depending on map size, the $^{12}$CO data cubes were obtained by
integrating for 0.6~sec or 1.2~sec (the ``sample time'') for each
7.3\arcsec\ pixel on the sky. A reference position (usually observed
at the end of each row in a scan), located at $4^h 12^m 24.9^s$
(J2000.0) 26\dg 50\arcmin\ 39\arcsec , was used throughout. A sample
time of 1.0~sec was used with the smaller $^{13}$CO/C$^{18}$O
maps.

The CO 3-2 data were observed in grade 3 weather ($\tau$(225~GHz) $
\sim$ 0.08--0.12); better weather ($\tau$(225~GHz) $\sim$ 0.05--0.065)  
was used for the $^{13}$CO 3-2 and C$^{18}$O 3-2 data. System
temperatures for the $^{12}$CO data ranged from 340 to 580~K ($<$450~K
in most regions); for the isotopologues temperatures were typically
300-500~K.

The regions mapped in $^{12}$CO are listed in Table~\ref{obs}; these
were defined based on large-scale extinction maps \citep{dob05}, the
surveys of dense cores conducted by OMK02, and the locations of
Herbig-Haro (HH) objects.  Only regions with bright $^{12}$CO emission
were mapped in $^{13}$CO and C$^{18}$O: two overlapping maps centred
near \hco\ cores 6, 7 and 11 were repeated a number of times to reduce
the noise in the final cube; the full map covers an area of
$\sim$22\arcmin $\times$12\arcmin , although note that the noise is
roughly a factor of two higher in the north and west portion
($\sim$60\%) of the map.  Similarly, two overlapping maps covering
cores 13a, 13b and 14 were repeated in the L~1495~SE ridge; in this
case, the map covers a rectangular area of approximately 20\arcmin
$\times$6\arcmin , though again the north-west quarter of the map is
about 3-times noisier than the rest of the data cube.  In all, eleven
overlapping maps were co-added in L~1495~E, and eleven in
L~1495~SE. The regions covered in $^{13}$CO and C$^{18}$O
are marked in Fig.~\ref{over}.

%%%%%%%%%%%%%%%%%%%%%%%%%%%
%%% Figure  %%%%%%%%%%%%%%%
\begin{figure}
\centering
\epsfxsize=8cm
%\epsfbox{/export/data/cdavis/jcmt/TAURUS/PAPER-FIGURES/Chan-E.eps}
\epsfbox{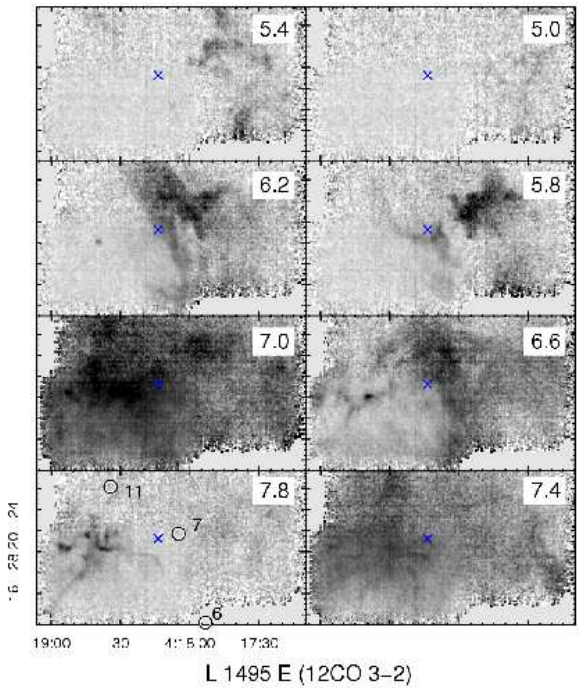}
\caption[] {$^{13}$CO 3-2 channel maps across L~1495~E, plotted between
4.8 and 8.0~\kms , in 0.4~\kms\ intervals.
\hco\ cores are marked and labelled bottom-left; the small cross in the 
centre of each map is drawn as a reference marker. The velocity
of the channel is marked top-right in each panel.}   
\label{chan-e}
\end{figure}
%%%%%%%%%%%%%%%%%%%%%%%%%%%

%%%%%%%%%%%%%%%%%%%%%%%%%%%
%%% Figure  %%%%%%%%%%%%%%%
\begin{figure}
\centering
\epsfxsize=8cm
%\epsfbox{/export/data/cdavis/jcmt/TAURUS/PAPER-FIGURES/Chan-SE.eps}
\epsfbox{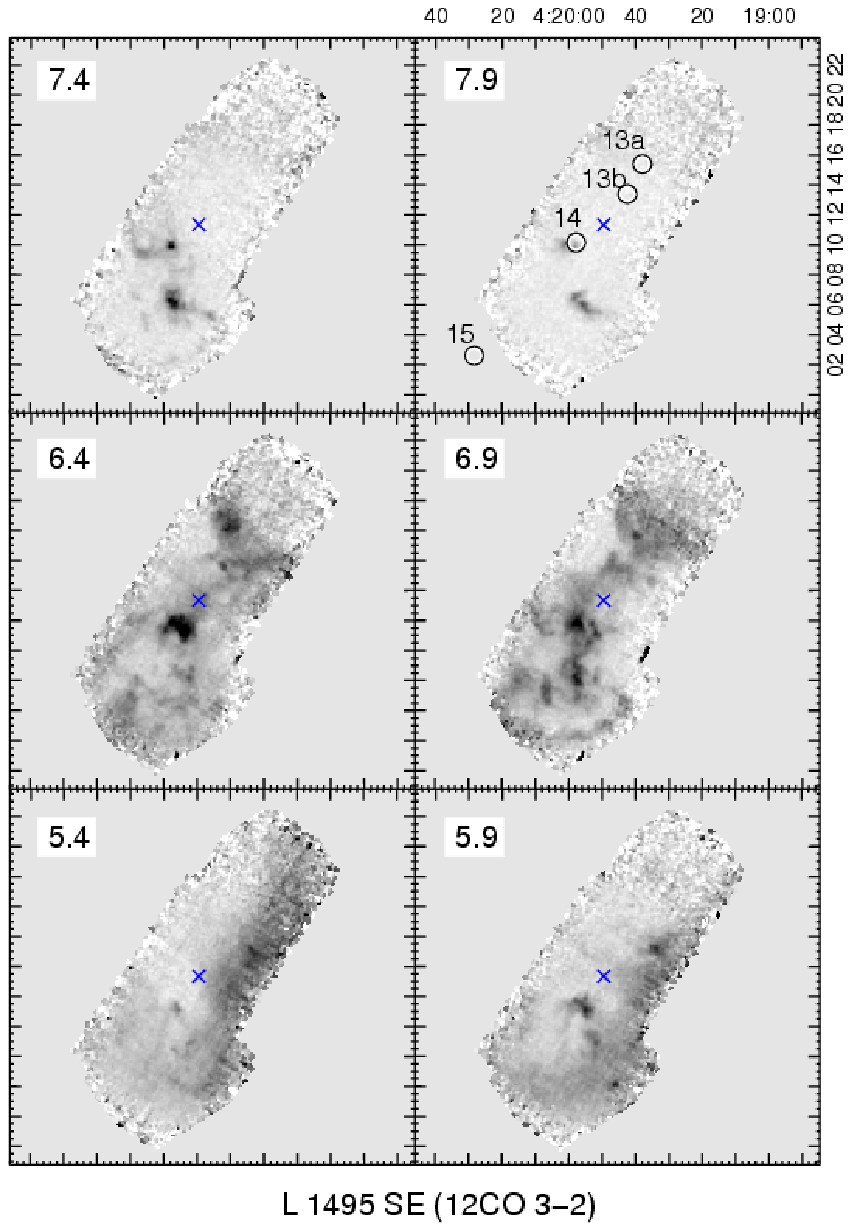}
\caption[] {$^{13}$CO 3-2 channel maps across L~1495~SE, plotted between
5.15 and 8.15~\kms , in 0.5~\kms\ intervals.
\hco\ cores are marked and labelled top-right; the small cross in the 
centre of each map is drawn as a reference marker.The velocity
of the channel is marked top-left in each panel.}   
\label{chan-se}
\end{figure}
%%%%%%%%%%%%%%%%%%%%%%%%%%%

%%%%%%%%%%%%%%%%%%%%%%%%%%%%%%%%%%%%%%%%%%%%%%%%%%%%%%%%%%%%%%%%%
%%%%%%%%%%%%%%%%%%%%%%%%%%%%%%%%%%%%%%%%%%%%%%%%%%%%%%%%%%%%%%%%%

\subsection{Data processing}

The individual CO 3-2 data cubes listed in Table~\ref{obs} (third
column) were reduced using the automated ORAC-DR pipeline
\citep{cav08}.  Bad or overly noisy spectra were identified by the  
pipeline, based on quality assurance criteria\footnote{Receiver system
temperature $>$600~K, noise across the spectrum exceeds
50\% of the mean noise level in all spectra, or
spectrum baseline deviates by more than 10\% from the
mean.} stipulated for the Gould Belt Survey legacy data, and these
were first removed and noisy regions at the band edges trimmed. The
resulting time-series data were re-gridded to a 3D data cube and
regions of line emission identified and masked out. A fifth-order
polynomial was fit to the resulting emission-free regions and
subtracted from each reduced cube. Cubes observed at the same spatial
and spectral position were combined and a new baseline region mask was
determined from this higher signal-to-noise cube. This was applied to
the individual cubes, which were then mosaicked and co-added together
using a light Gaussian smoothing kernel with a Full Width Half Maximum
(FWHM) $\sim$7.3\arcsec , to produce the final cubes and images. The
inverse of this baseline mask identifies the line emission regions and
this was used to determine the integrated intensity and velocity
images presented in this paper.  Note that the Gaussian smoothing
results in a spatial resolution to point sources of about
16\arcsec .

Special care was taken to remove striping from the data cubes when
present \citep{cur10}. This was done by determining
the receptor-to-receptor response for the whole time-series, and
dividing that response out. This technique does assume {\em a priori}
that each receptor is exposed to similar emission during the
scan. Given the diffuse and extended nature of the CO emission from
Taurus, this technique worked well in eliminating or, at the very
least, in minimising the striping.

In CO 3-2 the noise is relatively uniform across the extensive region
mapped. In the un-binned data (at the full 0.05\kms\ resolution), at
velocities within a few \kms\ of the ambient, the root-mean-square
(RMS) noise level ($T_A^*(rms)$) measures 0.10--0.20~K in L~1495-E, N
and S (being highest in the south), 0.08--0.15~K in L~1495-W (highest
in the north-west), and 0.08--0.16~K along the south-east ridge (it is
a little higher, $\sim$0.22~K, in the data around cores 10, 10b and
12a).

The $^{13}$CO and C$^{18}$O data were reduced in the same way. 
As noted above, the maps in $^{13}$CO and C$^{18}$O were repeated a
number of times.  We found that in order to produce the optimal coadded
map, the worst quality data had to be discarded before
being included in the map reconstruction. This was done by measuring
the RMS of each of the spectra in the input data and masking those
with the highest RMS before generating the map. An iterative method
was used to determine the optimal RMS value to apply as a mask. In
each step of the iteration, the data were masked using a different
limiting RMS before being reconstructed into a map using the Starlink
MAKECUBE routine \citep{jen08}. The RMS in the reconstructed map was
then measured. The limiting RMS was varied to produce a reconstructed
map with the lowest RMS. This map was taken to be the optimal map. In
the final, coadded maps the RMS noise level measures
$T_A^*(rms)\sim$0.13-0.20~K in $^{13}$CO and $\sim$0.16-0.25~K in
C$^{18}$O at the unbinned spectral resolution of 0.05~\kms.  As
mentioned earlier, the noise level is somewhat higher in the
north-west regions of each map.

In this article all images are presented in units of antenna temperature
($T^*_A$).  When calculating outflow parameters, integrated antenna
temperature has been converted to main beam brightness temperature,
$T_b = T^*_A/(\eta_{mb})$, using an aperture efficiency $\eta_{mb} =
0.63$ \citep{buc09}.

%%%%%%%%%%%%%%%%%%%%%%%%%%%%%%%%%%%%%%%%%%%%%%%%%%%%%%%%%%%%%%%%%
%%%%%%%%%%%%%%%%%%%%%%%%%%%%%%%%%%%%%%%%%%%%%%%%%%%%%%%%%%%%%%%%%

\section{Results}

\subsection{Overall distribution of CO 3-2 emission and large-scale
velocity structure}

The entire region mapped in CO 3-2 with HARP is shown in
Fig.~\ref{over}, where we also show integrated $^{13}$CO 3-2 images of the
L~1495-E and L~1495~SE regions observed in this isotopologue (and in
C$^{18}$O -- data not shown).  For comparison purposes, we also present
the $^{13}$CO 1-0 map of \citet{gol08}.

%%%%%%%%%%%%%%%%%%%%%%%%%%%
%%% Figure  %%%%%%%%%%%%%%%
\begin{figure}
\centering
\epsfxsize=8.0cm
%\epsfbox{/export/data/cdavis/jcmt/TAURUS/PAPER-FIGURES/iwc-e-lab.eps}
\epsfbox{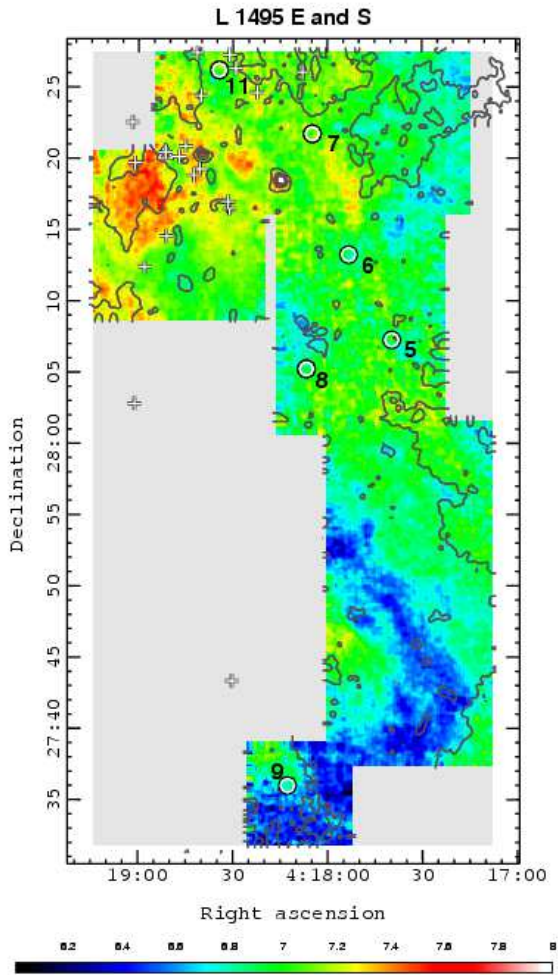}
\caption[] {A plot of intensity-weighted radial velocity (first moment) 
showing the distribution of blue and red-shifted CO 3-2 emission
around L~1495~E and L~1495~S. Black/blue through to red/white display
velocities in the range 6.0 to 8.0~\kms\ (green/yellow represent
intermediate velocities).  Circles, crosses and triangles mark the
positions of \hco\ cores (labelled), YSOs and HH objects,
respectively. Integrated CO 3-2 emission (zeroth moment, from
Fig.~\ref{over}) is over-plotted as contours; levels measure
4,6,8,10,12~K~\kms . Note that there is CO 3-2 line emission throughout 
this entire map.}
\label{radvel-e}
\end{figure}
%%%%%%%%%%%%%%%%%%%%%%%%%%%

When mapping L~1495 in CO 3-2 we followed the high-extinction ridge of
dense cores observed in C$^{18}$O 1-0 and \hco\ 1-0 by \citet{oni96}
and OMK02, respectively; the ridge is evident in the $^{13}$CO 1-0
integrated intensity map reproduced in Fig.~\ref{over}.  In
L~1495 the CO 3-2 emission is generally faint, diffuse and/or
optically thick: the integrated antenna temperature is $< 10.0$~K over
99\% of the mapped region, while we estimate the $^{12}$CO opacity,
$\tau_{12}$, to be in the range 3--38 in regions where $^{13}$CO has
been observed, (although values will be overestimated where profiles
are self-absorbed - see Section 3.2 for details). There are a number of
compact knots superimposed on this diffuse emission, most of which are
associated with outflows rather than cores.  Indeed, the CO 3-2
emission features are by-and-large unrelated to the compact cores
identified in \hco .  Unfortunately there is very little submm
continuum data in L~1495 for comparison purposes (in their pointed
1.3~mm survey of young stars, Motte \& Andr\'e [2001] observe only
nine sources in L~1495); only observations of cores 13a, 13b and 14
were found in the SCUBA (Submillimetre Common User Bolometer Array)
Legacy Catalogue \citep{fra08} and only the dense cores themselves
were detected. However, the entire area will be mapped at 450\mic\ and
850\mic\ with SCUBA-2 \citep{hol06} as part of the Gould Belt survey
\citep{war07a}. 

The $^{13}$CO and C$^{18}$O emission in the L~1495~E and L~1495~SE
regions mapped is very weak; in $^{13}$CO the peak intensity towards
the compact features seen in Fig.~\ref{over} measures only $T_A^* \le
2.0$~K and $\le 2.5$~K, respectively.  Profiles are generally narrow
($\sim$2~\kms\ wide) and are either flat-topped or double-peaked (and
therefore may in some regions be self-absorbed).  $^{13}$CO channel
maps for both regions are shown in Figs.~\ref{chan-e} and
\ref{chan-se}.  Note the small group of compact, marginally
red-shifted knots to the east of cores 7 and 11 in the L~1495-E
region, and the more diffuse, slightly blue-shifted emission to the
west and north-west.  Compact features are also evident along the
south-east ridge, at both blue and red-shifted velocities. In both
regions, however, with the possible exception of core 14 in
Fig.~\ref{chan-se}, we again see no obvious correlation with the \hco\
cores.

%%%%%%%%%%%%%%%%%%%%%%%%%%%
%%% Figure %%%%%%%%%%%%%%%%
\begin{figure*}
\centering
\epsfxsize=13.0cm
%\epsfbox{/export/data/cdavis/jcmt/TAURUS/PAPER-FIGURES/iwc-w-lab.eps}
\epsfbox{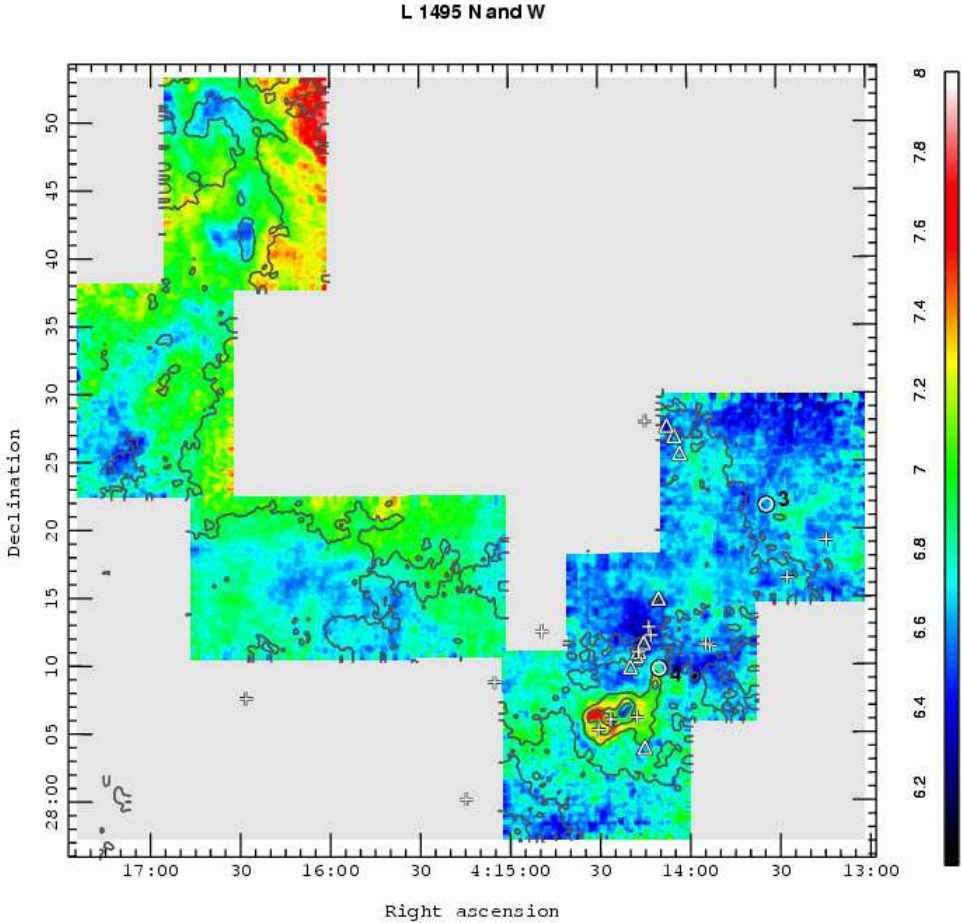}
\caption[] {Same as Fig.~\ref{radvel-e} but for the region labelled 
L~1495~N and L~1495~W in Fig.~\ref{over}. Again, there is CO 3-2 line
emission throughout this region. }
\label{radvel-w}
\end{figure*}
%%%%%%%%%%%%%%%%%%%%%%%%%%%

%%%%%%%%%%%%%%%%%%%%%%%%%%%
%%% Figure %%%%%%%%%%%%%%%%
\begin{figure*}
\centering
\epsfxsize=16.0cm
%\epsfbox{/export/data/cdavis/jcmt/TAURUS/PAPER-FIGURES/iwc-se-lab.eps}
\epsfbox{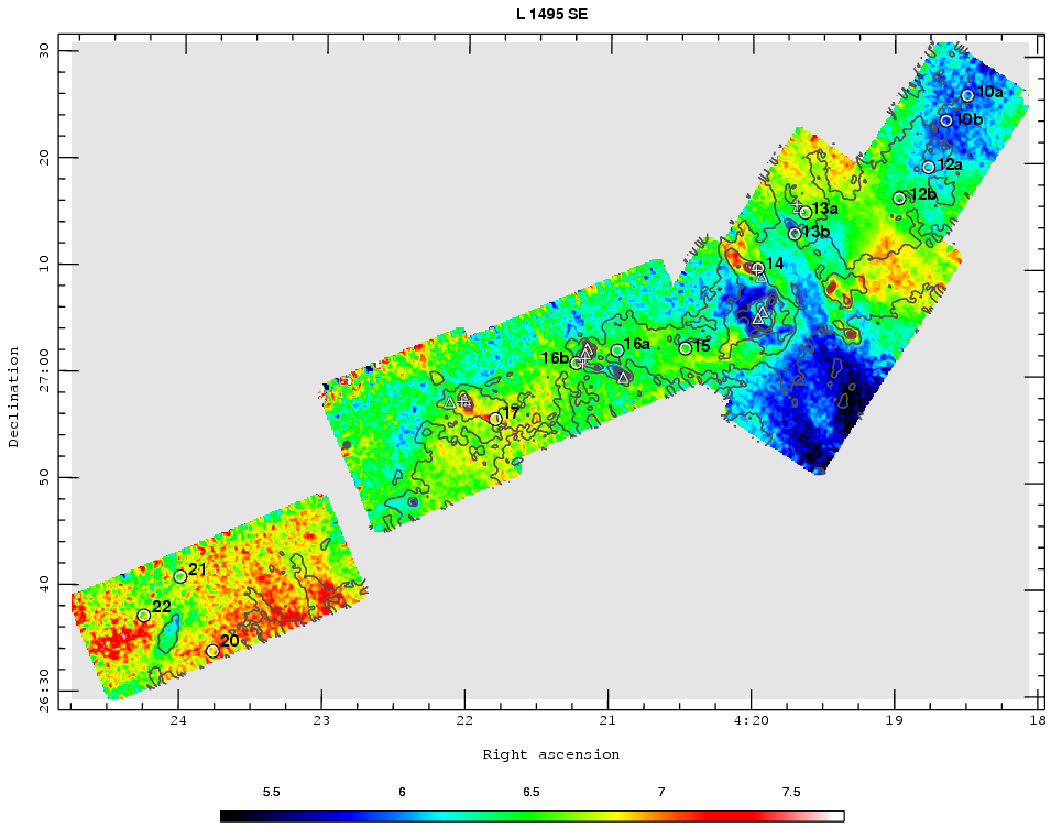}
\caption[] {Same as Fig.~\ref{radvel-e} but for the SE ridge in 
L~1495. In this region the colours span 5.3~\kms\ (black/blue) to
7.7~\kms\ (red/white). Once again, we detect CO 3-2 line emission
throughout this entire region.}
\label{radvel-se}
\end{figure*}
%%%%%%%%%%%%%%%%%%%%%%%%%%%

%%%%%%%%%%%%%%%%%%%%%%%%%%%%%%%%%%%%%%%%%%%%%%%%%%%%%%%%%%%%%%%%%
%%%%%%%%%%%%%%%%%%%%%%%%%%%%%%%%%%%%%%%%%%%%%%%%%%%%%%%%%%%%%%%%%

To illustrate the velocity structure observed in CO 3-2, we present in
Figs.~\ref{radvel-e}--\ref{radvel-se} intensity-weighted radial
velocity maps, colour-coded to show subtle changes
in the centroid velocity of the gas (note that CO 3-2 line emission is
detected at all locations in each image).  These reveal the collimated
blue- and red-shifted lobes of a number of outflows in L~1495, but
also large-scale velocity gradients and ``bubbles'' of gas associated
with clusters of young stars. In L~1495~E, for example, diffuse,
red-shifted emission envelopes the cluster of YSOs to the east of the
\hco\ cores 5-8 (Fig.~\ref{radvel-e}; the YSOs are marked with crosses
in this figure).  CO velocities to the west and south of this chain of
cores are more blue-shifted (a result confirmed by the CO 1-0
observations of Goldsmith et al. 2008).  The cluster of YSOs and the
associated gas may be detaching itself from the rest of the cloud;
alternatively, the red-shifted CO may represent a bubble being driven
eastward by the cluster of young stars.

In L~1495~W, \hco\ core 4 seems to be surrounded by a number of young
stars; those to the north coincide with a region of blue-shifted CO,
while those to the south coincide with a very distinct cloud of
red-shifted gas (Fig.~\ref{radvel-w}).  T~Tauri stars are often more
widely distributed than their younger protostellar counterparts and
prestellar cores; in L~1495~E and L~1495~W we may therefore be
witnessing the expansion of the natal cloud and the dispersal of these
young stars.

We also note the east-west velocity gradient seen in the L~1495~N
region.  The map in Fig.~\ref{radvel-w} suggests that the diffuse
background emission to the north-west of the L~1495 bowl -- in the
unobserved regions in the centre and top of the figure -- is likely to
be red-shifted with respect to the denser material in the bowl.
Again, this is evident in the CO 1-0 observations of \citet{gol08}.

Along the SE ridge, a number of collimated outflow lobes (discussed
further below) are superimposed onto a ridge of diffuse CO. In the
centre of Fig.~\ref{radvel-se} there is some indication that the
south-west edge of the ridge is red-shifted with respect to the
north-east edge.  The cores and YSOs in this area lie predominantly
along the boundary between this blue- and red-shifted gas.  If the
cores and young stars delineate the high-density axis of the L~1495
ridge, then this distribution of blue-shifted and red-shifted gas
could be symptomatic of rotation.

%%%%%%%%%%%%%%%%%%%%%%%%%%%
%%% Figure %%%%%%%%%%%%%%%%
\begin{figure}
\centering
\hspace*{-0.6cm}
\epsfxsize=9.4cm
%\epsfbox{/export/data/cdavis/jcmt/TAURUS/PAPER-FIGURES/AvSpec.eps}
\epsfbox{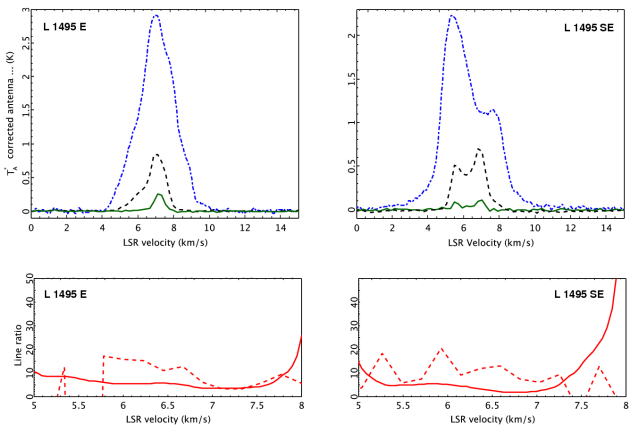}
\caption[] {{\em Top: } $^{12}$CO 3-2 (blue/dot-dashed), $^{13}$CO 3-2
(black/dashed), and C$^{18}$O 3-2 (green/full) spectra, averaged over
the full extent of the areas mapped simultaneously in $^{13}$CO and
C$^{18}$O in L 1495~E and L~1495~SE. The $^{13}$CO and C$^{18}$O
spectra have been binned to a resolution of 0.25~\kms\ ; the
resolution of the $^{12}$CO 3-2 spectra is 0.05~\kms . {\em Bottom: }
ratio of these averaged spectra; $^{12}$CO/$^{13}$CO full red lines,
$^{13}$CO/C$^{18}$O dashed red lines.}
\label{spec}
\end{figure}
%%%%%%%%%%%%%%%%%%%%%%%%%%%

%%%%%%%%%%%%%%%%%%%%%%%%%%%
%%% Figure %%%%%%%%%%%%%%%%
\begin{figure*}
\centering
\epsfxsize=14.0cm
%\epsfbox{/export/data/cdavis/jcmt/TAURUS/PAPER-FIGURES/PV-diags-all.eps}
\epsfbox{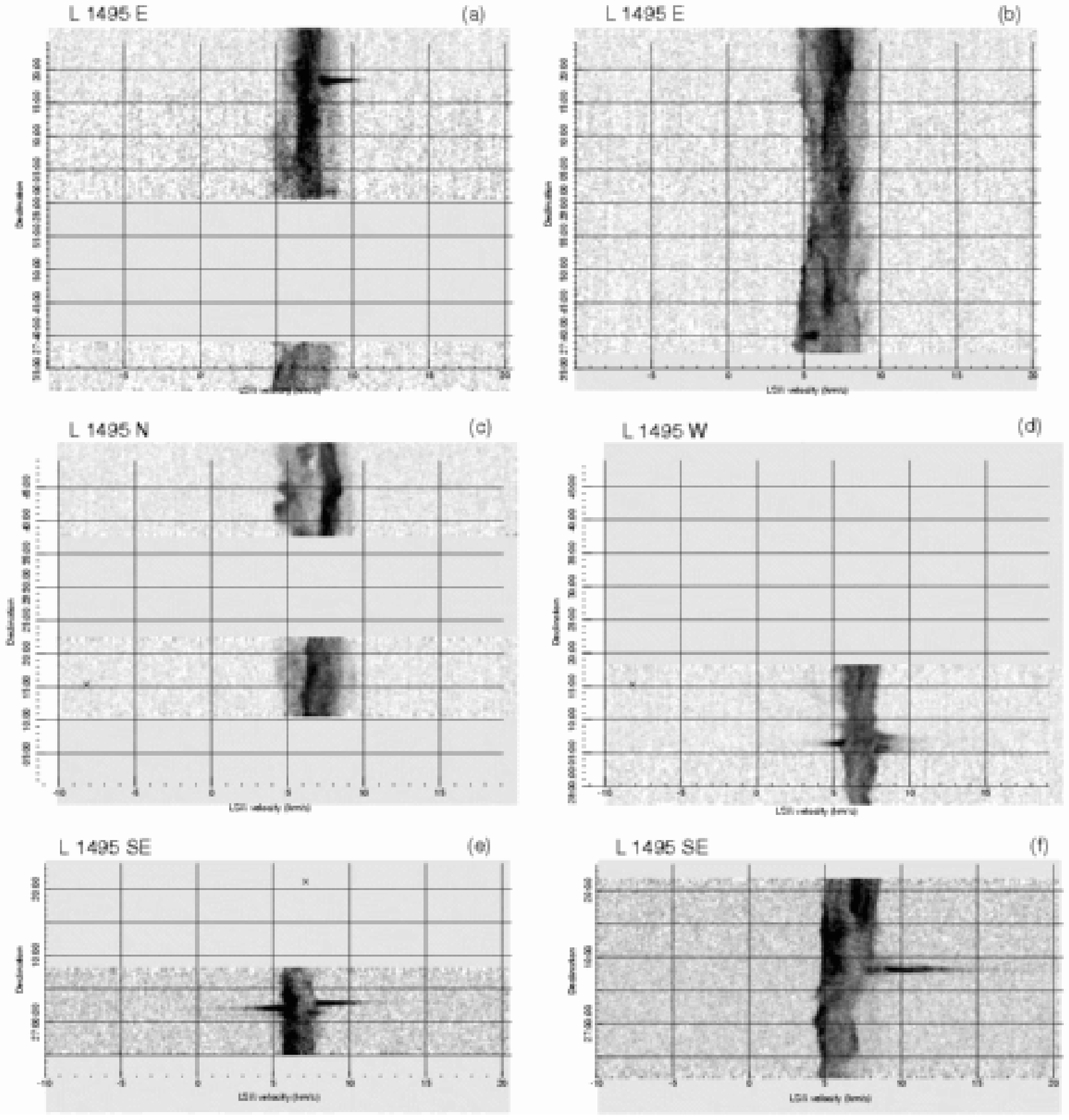}
\caption[] {CO 3-2 position-velocity diagrams measured along vertical 
slices through
L~1495~E  at RA 4:18:14.2 (a) and RA 4:17:30.1 (b), through
L~1495~N  at RA 4:16:30.2 (c),                      through 
L~1494~W  at RA 4:14:22.2 (d),                      and through
L~1495~SE at RA 4:21:11.1 (e) and RA 4:19:28.7 (f).}
\label{pv}
\end{figure*}
%%%%%%%%%%%%%%%%%%%%%%%%%%%

%%%%%%%%%%%%%%%%%%%%%%%%%%%%%%%%%%%%%%%%%%%%%%%%%%%%%%%%%%%%%%%%%
%%%%%%%%%%%%%%%%%%%%%%%%%%%%%%%%%%%%%%%%%%%%%%%%%%%%%%%%%%%%%%%%%

\subsection{Averaged spectra and PV diagrams}

Averaged spectra and Position-Velocity (PV) diagrams are shown in
Figs.~\ref{spec} and \ref{pv}.  The former illustrate the relatively
quiescent nature of the bulk of the gas in L~1495. They also give an
indication of the mean opacity in the three lines observed. If the gas
is optically thin then the line intensity ratios at a given velocity
should approach the abundance ratios, X[$^{12}$CO/$^{13}$CO]=70 and
X[$^{13}$CO/C$^{18}$O]=8.4 \citep{fre82,wil99}, provided that the
excitation temperature and beam efficiency and filling factors are the
same for all three lines.  Photo-dissociation and chemical
fractionation effects in low-extinction regions can also lead to
enhanced isotopic ratios \citep{lan89}.  At intermediate velocities
(5-8~\kms ), where line emission is detected in all three
isotopologues, the $^{13}$CO/C$^{18}$O line ratio is indeed close to
this canonical value (see lower panels in Fig.~\ref{spec}). However,
at these velocities the $^{12}$CO/$^{13}$CO ratio is of the order of
5-10, so the $^{12}$CO emission is clearly optically thick. 
For a $^{12}$CO/$^{13}$CO abundance ratio of 70, 
the $^{12}$CO opacity, $\tau_{12}$, is given by: 

\begin{equation}
      \frac {\int T^*_A(^{12}{\rm CO}) . \delta v} 
            {\int T^*_A(^{13}{\rm CO}) . \delta v} 
             = 
      \frac {( 1-e^{-\tau_{12}} )} {( 1-e^{-\tau_{13}} )}
%            \sim
%     \frac {70( 1-e^{-\tau_{12}} )} {\tau_{12}}
\end{equation}

%The approximation assumes that $\tau_{13} << 1$.
where $\tau_{13} = \tau_{12}/70$.
A $^{12}$CO/$^{13}$CO line ratio of $\sim 10$ thus results in
$\tau_{12} \sim 7.4$. Again, this value will be overestimated since
$^{12}$CO is almost certainly self-absorbed; a lower abundance ratio
\citep[e.g.][]{gol08} would also result in a lower opacity.  The ratio
does increase towards blue-shifted and, particularly, red-shifted
velocities. It therefore seems likely that, in the extended regions
where the bulk of the outflowing gas is found, the high velocity line
emission used to calculate the outflow parameters listed below will be
only marginally optically thick.

PV diagrams, some of which are shown in Fig.~\ref{pv}, are useful for
distinguishing compact molecular outflows from turbulent motions in
the large-scale cloud. The outflows are seen as discrete horizontal
spikes; turbulence appears as regions of diffuse emission that are
more extensive along the spatial axis (vertically in these plots)
though confined to low velocities, i.e. to within a few \kms\ of the
ambient gas velocity.  Indeed, we suggest that ``eyeballing'' outflows
in PV diagrams is a relatively reliable way of finding them in HARP
data cubes, since they are spatially (in one axis at least) as well as
spectrally distinct from turbulent cloud motions and multiple cores
seen along the same sight-line.

We have therefore searched for outflows by stepping through our data
cubes in latitude and then in longitude, identifying spikes in PV
space that extend more than $\pm$2.5~\kms\ from the local ambient
velocity with an intensity of $>$0.3~K (roughly the 2$\sigma$
level). A spike is defined as being narrower than 1\arcmin\ in the
spatial direction (i.e. in longitude and/or in latitude); we adopt
1\arcmin\ as a conservative upper limit to the width of an outflow
lobe.  Outflows are subsequently verified in red-shifted and
blue-shifted integrated intensity maps, most of which are presented in
the next section.

A number of outflows are evident in Figs.~\ref{pv}. In the L~1496~E
region, for example, a compact, red-shifted knot of emission
$\sim$8\arcmin\ to the west-south-west of CoKuTau-1, which we later
refer to as E-CO-R1, is very distinct at declination $\sim$28\dg
18\arcmin\ in Fig.~\ref{pv}a.  This feature is also evident
$\sim$5\arcmin\ south-east of core 7 in Fig.~\ref{radvel-e}. Similar,
though less extreme spikes extending red-ward and blue-ward of the
ambient gas are seen to the north of V892~Tau (not shown, though
discussed further below).  Fig.~\ref{pv}b is typical of PV diagrams
further west, toward the centre of the L~1495 bowl, where we find no
clear-cut examples of molecular outflows, though where the gas
generally appears more turbulent.  This is particularly true in the
region labelled L~1495~N in Fig.~\ref{over}, where the emission
profiles are double-peaked (top section in Fig.~\ref{pv}c).  The CO
3-2 emission appears somewhat more ordered to the south and west of
this region (see for example Fig.~\ref{pv}c [lower section] and
Fig.~\ref{pv}d).

Along the south-eastern ridge in L~1495, PV spikes associated with
high-velocity outflows are more prevalent.  Example PV diagrams, again
plotted along the declination axis, are presented in Fig.~\ref{pv}e
and \ref{pv}f: these show emission from the complex HH~392 region and
the red lobe of the collimated bipolar outflow associated with IRAS
04166+2706 and core 13b, respectively (discussed further below).

%%%%%%%%%%%%%%%%%%%%%%%%%%%%%%%%%%%%%%%%%%%%%%%%%%%%%%%%%%%%%%%%%%%%
%%%%%%%%%%%%%%%%%%%%%%%%%%%%%%%%%%%%%%%%%%%%%%%%%%%%%%%%%%%%%%%%%%%%

\subsection{Outflows in L~1495}

When trying to identify outflows in complex regions like L~1495 one
ideally needs a measure of the systemic velocity associated with each
outflow driving source.  This is difficult to do using the CO 3-2
observations alone (Fig.~\ref{pv}); even the $^{13}$CO and C$^{18}$O
profiles are double-peaked in some regions (Fig.~\ref{spec}).  We
therefore use the \hco\ observations of OMK02 as a guide; these data
are more likely to be optically thin and trace only the quiescent,
high-density gas in each area.  Cores are mapped throughout the bowl
and south-east ridge in L~1495. The \hco\ data also suggest that the
ambient velocity does not change drastically across the region
(Table~\ref{obs}), a result which is supported by our CO 3-2
observations (particularly when one scans through the CO 3-2 data
cubes in PV space), and our more optically thin, though
less-extensive, $^{13}$CO and C$^{18}$O data (Figs.~\ref{chan-e} and
\ref{chan-se}).

\begin{figure}%
\centering
\epsfxsize=8.3cm
%\epsfbox{/export/data/cdavis/jcmt/TAURUS/PAPER-FIGURES/L1495E-flows-lab.eps}
\epsfbox{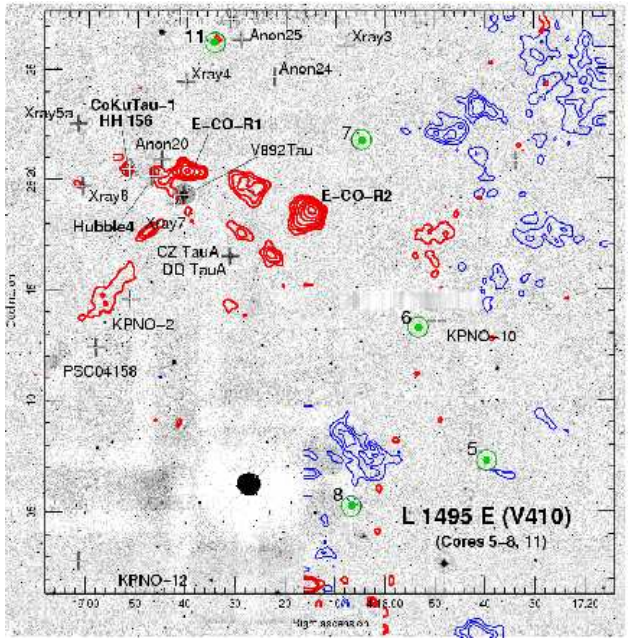}
\caption[] {High-velocity CO 3-2 emission in  L~1495~E, integrated
between 4.0 and 6.0\kms\ (blue/thin contours) and 8.0 and 10.0\kms\ (red/thick
contours), plotted as contours on top of a narrow-band H$_2$ ($+$ continuum) 
image of the region; contour levels are 1.75,2,2.5,3,4,5,7,10~K \kms .
Green circles, crosses and triangles mark \hco\ cores, YSOs and HH
objects, respectively.}
\label{co-e}
\end{figure}

\begin{figure*}%
\centering
\epsfxsize=17.0cm
%\epsfbox{/export/data/cdavis/jcmt/TAURUS/PAPER-FIGURES/L1495W-flows-lab.eps}
\epsfbox{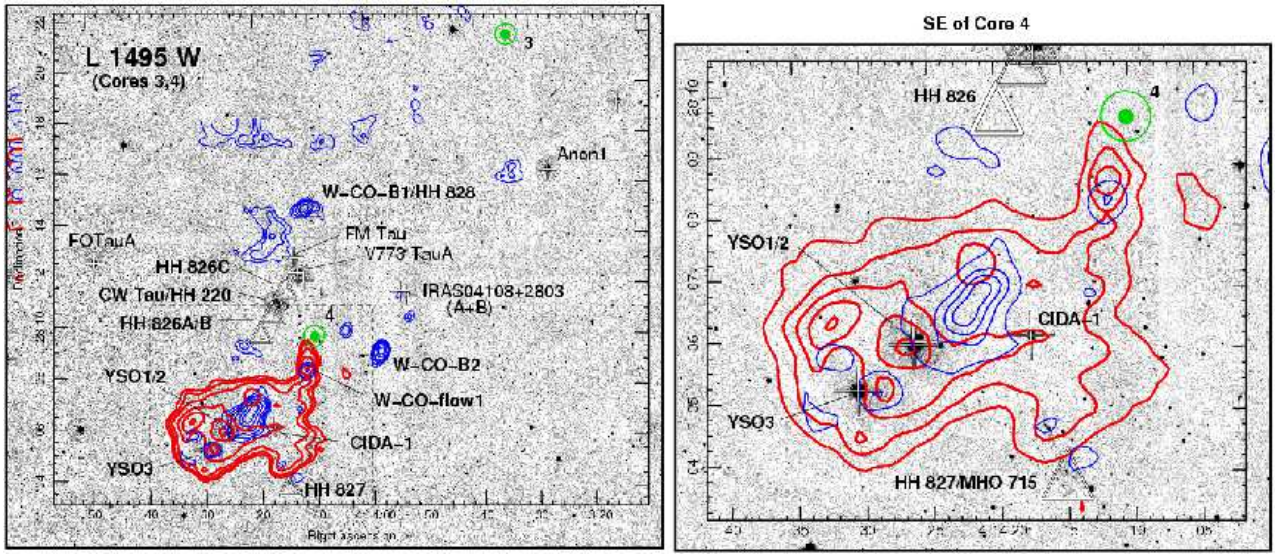}
\caption[] {Same as Fig.~\ref{co-e} but for the L~1495~W region.  {\em
Left -} high-velocity CO 3-2 integrated between 3.7 and 5.7\kms\
(blue/thin contours) and 7.7 and 9.7\kms\ (red/thick
contours). Contour levels are 1.0(blue-shifted
only),1.2,1.5,2,3,4,5,7,10~K \kms .  {\em Right -} a close-up view of
the region to the south-east of core 4 (marked with a dashed square at
left); contours measure 1,2,3,4,5,7,10~K \kms .}
\label{co-w}
\end{figure*}

When constructing integrated intensity maps of the high-velocity blue
and red-shifted gas, we therefore adopt systemic velocities of
7.0\kms , 6.7\kms\ and 6.6\kms\ for the L~1495~E, L~1495~W and
L~1495~SE regions, respectively. Contour maps for select regions are
presented in Figs.~\ref{co-e} - \ref{co-se4}, where we also over-plot
the positions of the Taurus YSOs, known HH objects and \hco\
cores for reference.  

In these figures we make use of shallow though extensive near-infrared
images of Taurus, obtained as part of the U.K. Infrared Deep Sky
Survey currently being conducted at the United Kingdom Infrared
Telescope \citep[UKIDSS:][]{law07}.  Images in broad-band K and
narrow-band \htwo\ emission (hereafter referred to simply as H$_2$)
have been secured for the entire Taurus-Auriga-Perseus complex.  The
wide-field camera (WFCAM) used to obtain these data, the data
reduction procedure, and the WFCAM archive are described in detail by
\citet{luc08} and \citet{dav08}.  Note that we also use the
acronym ``MHO'' for the molecular hydrogen emission-line objects --
the shock-excited infrared counterparts to HH objects -- observed in
L~1495\footnote{The acronym MHO has recently been approved by the
International Astronomical Union (IAU) Working Group on Designations
and has been entered into the Dictionary of Nomenclature of Celestial
Objects (http://cdsweb.u-strasbg.fr/cgi-bin/Dic?MHO)}. A complete
catalogue of all known galactic MHOs is available
on-line\footnote{http://www.jach.hawaii.edu/UKIRT/MHCat/} and 
in \citet{dav10}.

Below we discuss the star forming regions in L~1495 individually.

%%%%%%%%%%%%%%%%%%%%%%%%%%%
%%% Figure %%%%%%%%%%%%%%%%
%%%%%%%%%%%%%%%%%%%%%%%%%%%

%%%%%%%%%%%%%%%%%%%%%%%%%%%
%%% Figure %%%%%%%%%%%%%%%%
%%%%%%%%%%%%%%%%%%%%%%%%%%%

%%%%%%%%%%%%%%%%%%%%%%%%%%%%%%%%%%%%%%%%%%%%%%%%%%%%%%%%%%%%%%%%%%%%%%%%
%%%%%%%%%%%%%%%%%%%%%%%%%%%%%%%%%%%%%%%%%%%%%%%%%%%%%%%%%%%%%%%%%%%%%%%%

\subsubsection{L~1495~E: CoKu~Tau-1 and HH~156}

L~1495~E harbours a cluster of about a dozen TTSs, many of which have
been identified through X-ray observations \citep[the V~410 X-ray and
Anon\# sources;][]{str94}. OMK02 label five cores in this region. In
the integrated intensity maps in Fig.~\ref{over} the CO is diffuse and
optically thick across much of the region; only a few compact peaks
are detected, all of which are marginally red-shifted and unrelated to
the \hco\ cores (Fig.~\ref{co-e}, though note that the \hco\ map does
not cover the region south of the young star Xray~7 and east of core
8).

Two seemingly unrelated red-shifted knots are identified as molecular
outflow features, E-CO-R1 and E-CO-R2.  Neither is obviously related
to any of the YSOs or \hco\ cores; their progenitors remain unknown,
although candidate sources surround E-CO-R1, and this object does
coincide with a peak in our $^{13}$CO integrated intensity map in
Fig.~\ref{over}.  Note also that in the H$_2$ images we detect no
line-emission features in this region.

In Fig.~\ref{co-e} CoKu~Tau-1 seems to be associated with a third
compact red-shifted CO feature. However, this low-velocity knot of
emission fails our search criteria for outflows (described above) and
so is not identified as such here.  CoKu~Tau-1 does power HH~156~A and
B, two compact HH objects offset $\sim$2\arcsec\ and
$\sim$12\arcsec\ to the south-south-west; the CO peak could represent
the counter-lobe, though deeper and/or higher-resolution CO data are
needed to be sure. (The lack of blue-shifted CO around HH~156 may be a
result of the HH flow exiting the cloud, or may be due to the
dispersal of the core around this TTS.)

Apart from CoKu~Tau-1, none of the young stars in the region are
obviously driving CO outflows; presumably most are too evolved.  Of
the dozen YSOs to the east of cores 5-8 in Fig.~\ref{co-e}, all bar
three have neutral {\em Spitzer}-IRAC colours ([3.6]-[4.5]$\sim$0;
[5.8]-[8.0]$\sim$0), consistent with relatively evolved (weak-line)
TTSs and an absence of mid-infrared excess due to a lack of
circumstellar material \citep[][note that V~892~Tau is
saturated]{luh06}. Only CoKu~Tau-1, CZ~TauA and DD~Tau~A have 2MASS
near-IR and {\em Spitzer} mid-IR magnitudes that brighten towards
longer wavelengths, consistent with them being protostars or very
young T~Tauri stars and therefore possible molecular outflow
candidates.  This low number of protostars in L~1495~E is consistent
with the paucity of outflows in our H$_2$ and CO data.

\subsubsection{L~1495~W: CW~Tau and HH~220/826-828}

The high-velocity CO in L~1495~W is shown in Fig.~\ref{co-w}.  Again,
the positions of candidate YSOs are marked, although note that the
{\em Spitzer} observations of \citet{luh06} do not cover this
region. The most dramatic feature in this area is the bubble of
red-shifted gas that surrounds a small group of young stars, YSOs
1,2,3\footnote{Kenyon et al. (2009) refer to these objects as MHO~1,
MHO~2 and MHO~3, using the acronym MHO for ``Mill House Observatory''.
We identify these sources with YSO so as to distinguish them from
Molecular Hydrogen emission line Objects (MHOs).} and CIDA-1 (see
also Fig.~\ref{radvel-w}).  More compact blue-shifted features are
also observed in L~1495~W.

At least four molecular outflows exist in this region: 
 (1) a compact north-south bipolar flow, W-CO-flow1, centred
     $\sim$1\arcmin\ south of (and seemingly unrelated to) \hco\
     core 4,
 (2) a collimated blue-shifted lobe extending north-westward from YSOs
     1 and 2 (which presumably also drive some [or all] of the
     red-shifted CO to the south-east),
 (3) a compact blue-shifted CO feature, W-CO-B1, associated with
     HH~828, and
 (4) a second blue CO knot,  W-CO-B2, to the south-west of core 4, which
     may well be driven by IRAS~04108+2803A; this CO knot is certainly
     extended towards this object, which is also rather red (J-H = 3.10, 
     H-K$_s$ = 2.32).

Two additional flows are detected in HH and H$_2$ emission,
HH~220/826, a jet from CW~Tau \citep{mcg04}, and an arc of HH/H$_2$
emission, HH~827/MHO~715, which may be associated with low-velocity
red-shifted CO extending southward from CIDA-1 (though again this
emission fails our outflow criteria).

YSO 1/2, YSO 3 and CIDA-1 have near-IR 2MASS colours consistent with
youth. The TTSs CW~Tau, V773~Tau~A and FM~Tau are more evolved and
appear to be associated with a completely separate region of star
formation.  CW~Tau is thought to be the most active classical TTS in
this region.  Even so, we detect no H$_2$ line emission from its HH
objects, and only HH~828 appears to be associated with high-velocity
(blue-shifted) CO emission.

%%%%%%%%%%%%%%%%%%%%%%%%%%%
%%% Figure %%%%%%%%%%%%%%%%
\begin{figure*}%
\centering
\epsfxsize=16cm
%\epsfbox{/export/data/cdavis/jcmt/TAURUS/PAPER-FIGURES/L1495SE3-flows-lab.eps}
\epsfbox{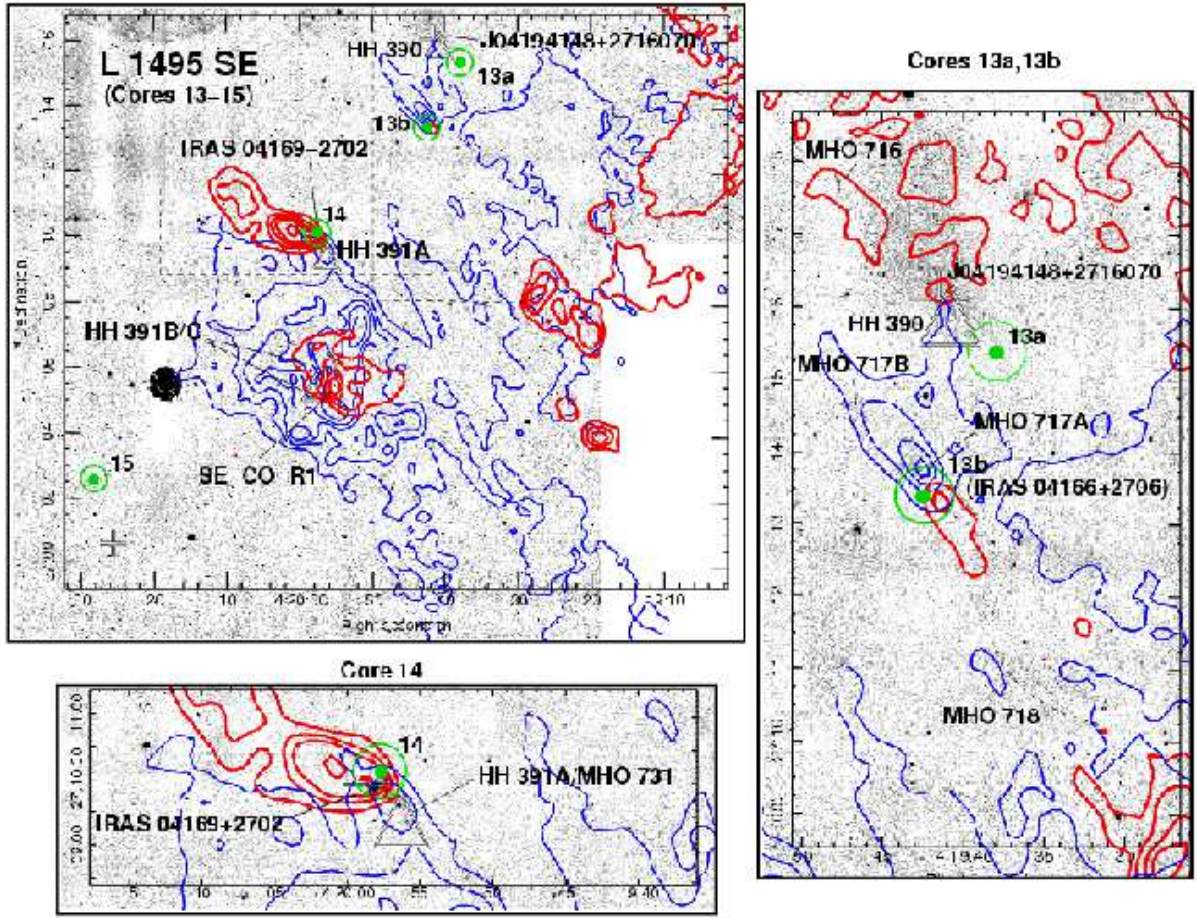}
\caption[] {Same as Fig.~\ref{co-e} but for the region around cores
13-15 in L~1495~SE. {\em Top-left -} high-velocity CO 3-2 integrated
between 3.6 and 5.6\kms\ (blue/thin contours) and 7.6 and 9.6\kms\
(red/thick contours); levels are 2,3,4,5,7,10~K \kms\ for both blue-
and red-shifted emission. {\em Right and Bottom-left -} close-up views
of the outflows associated with cores 13a,13b and 14: contour levels
are 1.5(red only),2,3,4,5,7,10~K \kms\ in the former and 2,3,4,7,10~K
\kms\ in the latter.}
\label{co-se3}
\end{figure*}

%%%%%%%%%%%%%%%%%%%%%%%%%%%%%%%%%%%%%%%%%%%%%%%%%%%%%%%%%%%%%%%%%%%%%%%%
%%%%%%%%%%%%%%%%%%%%%%%%%%%%%%%%%%%%%%%%%%%%%%%%%%%%%%%%%%%%%%%%%%%%%%%%

\subsubsection{L~1495~SE: HH~390-391}

The lower portion of the L~1495 ridge is populated with a number of
young stars and HH flows. The area around cores 13a, 13b, 14 and 15 is
shown in Fig.~\ref{co-se3}. Two of these four \hco\ cores (13b and
14) are associated with IRAS sources and HH objects. 

IRAS~04169+2702 (core 14) drives a collimated, possibly precessing
bipolar molecular outflow and HH object (HH~391A); the HH object is
also identified in H$_2$ emission, MHO~731 \citep[see also][]{gom97}.

A second, rather spectacular bipolar CO outflow emanates from core
13b.  This flow is particularly striking in Fig.~\ref{radvel-se}. The
source of the outflow, IRAS~04166+2706, is identified at submm
wavelengths \citep{san09}, though it does not appear in the 2MASS
point source catalogue nor in the YSO lists of \citet{luh06} and
\citet{ken09}.  It must therefore be particularly young, as expected
for molecular outflow and MHO progenitors (Hatchell et al. 2007b;
Davis et al. 2008, 2009).  Very faint H$_2$ features, MHO~717~A/B and
MHO~718, are observed in both flow lobes.  This outflow is well-known
from previous CO observations: \citet{taf04} report the discovery of a
collimated, Extremely High Velocity (EHV) outflow from IRAS~04166+2706, which
\citet{san09} have since observed in SiO 2-1 emission.  Both groups present
CO 2-1 spectra with striking peaks in the line wings at -35~\kms\ and
+45~\kms .  However, this EHV outflow component is only marginally
detected (at the 2-3$\sigma$ level) in our 3-2 data.  Based on a noise
level of $T^*_A(rms) \sim 0.1$~K in our data (equivalent to a main
beam brightness temperature of $\sim 0.2$~K), we set an upper limit of
$\sim$0.5 for the 3-2/2-1 intensity ratio (note that the beam sizes
are comparable).  This ratio suggests that the clumps or ``molecular
bullets'' associated with the EHV flow from IRAS~04166+2706 are
relatively cold and/or diffuse \citep[$n_{H2} \le 10^4$~cm$^{-3}$, $T
\le 50$~K;][]{yeh08}.

Core 13a drives a much less well defined molecular
outflow. MHO~716 may be associated with a red-shifted outflow lobe;
HH~390 \citep[which extends towards the south-west;][]{gom97} would
then be excited in the blue lobe.  The 2MASS source J04194148+2716070,
which \citet{luh06} and \citet{ken09} mis-identify as
IRAS~04166+2706, is situated $\sim$50\arcsec\ to the north-east of
core 13a, and is a possible driving source for this outflow.

%%%%%%%%%%%%%%%%%%%%%%%%%%%
%%% Figure %%%%%%%%%%%%%%%%
\begin{figure*}%
\centering
\epsfxsize=16cm
%\epsfbox{/export/data/cdavis/jcmt/TAURUS/PAPER-FIGURES/L1495SE4-flows-lab.eps}
\epsfbox{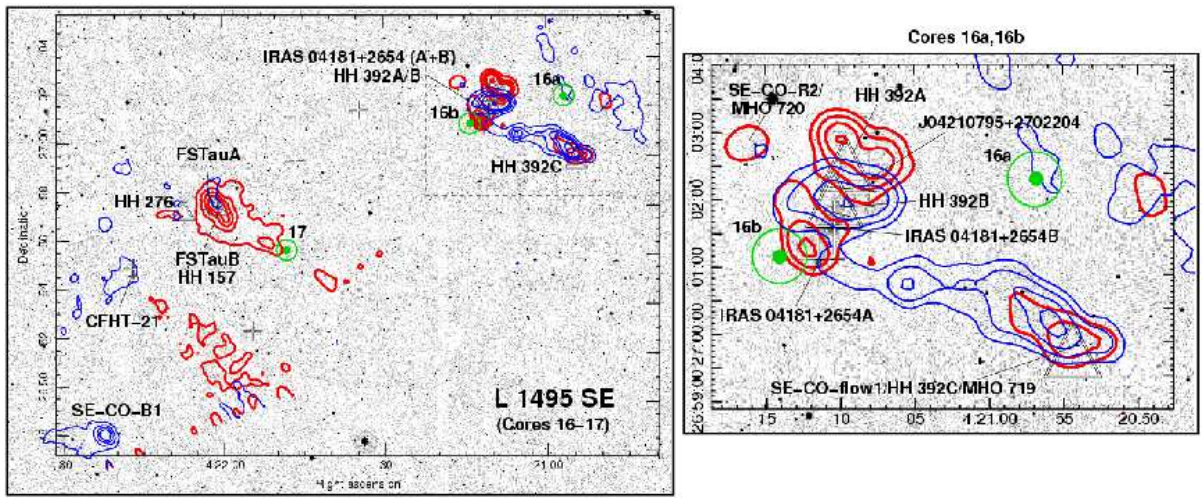}
\caption[] {Same as Fig.~\ref{co-e} but for the region around cores
16a,16b and 17 in L~1495~SE.  {\em Left -} high-velocity CO 3-2
integrated between 3.6 and 5.6\kms\ (blue/thin contours) and 7.6 and
9.6\kms\ (red/thick contours); levels are 1,2,3,4,5,7,10~K \kms\ for
both blue- and red-shifted emission. {\em Right -} a close-up of the
outflows associated with cores 16a and 16b; contour levels are
1,2,4,7,10~K \kms .}
\label{co-se4}
\end{figure*}
%%%%%%%%%%%%%%%%%%%%%%%%%%%%

Considering these three cores (13a, 13b and 14) and their associated
YSOs together, IRAS~04166+2706 is presumably the youngest of the
three; this is consistent with its powerful EHV CO outflow, its H$_2$
shocks, and the {\em absence} of HH objects.  IRAS~04169+2702 is
probably somewhat more evolved, having infrared colours that are
consistent with a Class I protostar (H-K$_s$ = 2.6; [3.6]-[5.8] =
2.1). J04194148+2716070 is bluer and is almost certainly a TTS, which
would explain its association with an HH object and the lack of
well-defined CO 3-2 outflow lobes.

Further south, HH~391~B and C represent a chain of optical features
that extend over $\sim$1\arcmin\ in a north-south direction
\citep{gom97}.  These objects coincide with large, complex bubbles of
both blue- and red-shifted emission.  HH~391~B/C are unlikely to be
associated with the IRAS~04169+2702/HH~391~A outflow. No H$_2$
emission was detected in this region, and the relationship between
these shock features and the complex CO emission in this area is
unclear, although high-velocity CO does coincide with these objects
(we label the more clearly-defined red-shifted emission SE-CO-R1).
Moreover, a $^{13}$CO peak does coincide with HH~391~B/C in our
integrated intensity map in Fig.~\ref{over}, which may be a sign of
youth.

\subsubsection{L~1495~SE: FS~Tau A/B (Haro 6-5B) and HH~157/276/392}

Moving further south-east along the L~1495 ridge, molecular outflows are
detected around IRAS~04181+2654 and the well-studied FS~Tau A/B
(Haro~6-5B) system (Fig.~\ref{co-se4}).  

IRAS~04181+2654~A is closest to core 16b and appears to drive a
bipolar CO outflow in a roughly east-west direction.  The
blue lobe of this flow may overlap with a second bipolar outflow,
SE-CO-flow1, which drives the compact shock features HH~392~C/MHO~719.
There is no source candidate for SE-CO-flow1 in the 2MASS or IRAS
point source catalogues, and \citet{luh06} and \citet{ken09} do not
identify a YSO near these objects. IRAS~04181+2654~A is resolved by
2MASS and exhibits an excess consistent with a protostar (J-H =
3.58, H-K$_s$ = 2.30).

IRAS~04181+2654~B lies $\sim$30\arcsec\ north-west of star A; this
nebulous infrared source is associated with an arc of HH emission,
HH~392~B (tentatively detected here in H$_2$) and also possesses
considerable excess in the near-IR (J-H = 5.11, H-K$_s$ = 2.70).  The
third source in the region, 2MASS source J04210795+2702204, has less
extreme near-IR colours (J-H = 1.8, H-K$_s$ = 1.5), though may
nonetheless also contribute to the complex high-velocity CO evident in
Fig.~\ref{co-se4}.  The ``kidney'' shape of the blue- and red-shifted
features coincident with HH~392~A and B certainly suggest the presence
of multiple flow lobes.  

Approximately 1.5\arcmin\ to the north-east of IRAS~04181+2654 there is a
faint knot of H$_2$ emission, MHO~720, which coincides with a knot of
red-shifted CO, SE-CO-R2, and a very marginal detection of
blue-shifted CO offset $\sim$15\arcsec\ to the south-west.  In all,
there are at least four molecular outflows within a 5\arcmin\ radius
of cores 16a and 16b.

FS~Tau A (J-H = 1.47, H-K$_s$ = 1.73) and FS~Tau~B (J-H = 1.06,
H-K$_s$ = 1.60; also known as Haro~6-5B) are a pair of TTSs, the
latter being associated with the spectacular HH jet and bow shock
HH~157 \citep{eis98}.  We detect red-shifted CO emission from this
outflow, though no H$_2$ line emission. HH~276 is a chain of faint HH
knots that crosses the HH~157 jet $\sim$1\arcmin\ to the north-east of
FS~Tau~B, although no high-velocity CO is clearly identified from this
flow.

Lastly, south-east of FS~Tau A, we note the very tentative detection
of a bipolar CO outflow from CFHT-21, orientated NE-SW, roughly
parallel with the FS~Tau~B jet.  This marginal detection is consistent
with the near-IR colours of this TTS (J-H = 1.5, H-K$_s$ =
1.0). SE-CO-B1 is a compact though massive high-velocity knot with a
tail extending along the south-east ridge.  This feature is, however,
not related to any known YSOs in the region.

%%%%%%%%%%%%%%%%%%%%%%%%%%%
%%% Table  %%%%%%%%%%%%%%%%

\begin{table*}
\centering
\begin{minipage}{195mm}
\begin{tiny}

\caption{Parameters for the outflows in L~1495. The top nine flows are in the bowl; 
the remaining flows are in the south-east ridge.}
\begin{tabular}{@{}lccc ccl ccccc@{}} 

 \hline
YSO or Outflow$^a$&Area$^a$&YSO      & RA$^a$  & Dec $^a$&CO flow$^b$&HH/MHO$^c$&\hco&$M$$^d$ &$|v_o-v|$$^d$&$P^d$&$E^d$\\
                  &        & type$^a$&(J2000.0)&(J2000.0)&   &  &core$^c$&(\Msol )&\kms  &(\Msol &($\times 10^{34}$\\
          &        &        &         &         &           &          &        &        &     & \kms) &    J)           \\
\hline

%%CoKu~Tau-1        & E & TTS        & 4:18:51.5 & 28:20:26 & Red     & HH156         &     & 0.0035 & 2.5 & 0.009 &  2 \\  
CoKu~Tau-1          & E & TTS        & 4:18:51.5 & 28:20:26 & No?     & HH156         &     &   --   & --  &  --   & -- \\ 
E-CO-R1$^e$         & E & --         & 4:18:39.5 & 20:20:20 & Red     &               &     & 0.0098 & 2.3 & 0.022 &  5 \\
E-CO-R2$^e$         & E & --         & 4:18:15.0 & 28:18:30 & Red     &               &     & 0.0256 & 4.0 & 0.102 & 41 \\
 \hline

%W-CO-R1            & W & --         & 4:14:31.1 & 28:06:30 & Red     &               &     & \\
%YSO~1/2            & W & Class I    & 4:14:26.3 & 28:06:03 & Blue    &               &     & 0.0201 & 4.2 & 0.084 & 35 \\ 
YSO~1/2             & W & Class I    & 4:14:26.3 & 28:06:03 & Bipolar &               &     & 0.0156 & 4.7 & 0.074 & 35 \\ 
CIDA-1              & W & TTS        & 4:14:17.6 & 28:06:10 & No?     & HH827/MHO715  &     &   --   & --  &  --   & -- \\
CW~Tau              & W & TTS        & 4:14:17.0 & 28:10:58 & No?     & HH220/826     &     &   --   & --  &  --   & -- \\ 
W-CO-B1$^e$         & W & --         & 4:14:10.7 & 28:14:40 & Blue    & HH828         &     & 0.0028 & 3.1 & 0.009 &  3 \\ 
W-CO-flow1$^e$      & W & --         & 4:14:12.0 & 28:08:30 & Bipolar &               & 4?  & 0.0038 & 3.1 & 0.012 &  4 \\ 
IRAS~04108+2803A?   & W & Class I    & 4:13:57.4 & 28:09:10 & Blue    &               &     & 0.0055 & 3.3 & 0.018 &  6 \\
 \hline

J04194148+2716070   & SE & TTS        & 4:19:41.5 & 27:16:07 & No      & HH390/MHO716 & 13a?&   --   & --  &  --   & --  \\ 
IRAS~04166+2706$^f$ & SE & Class 0?   & 4:19:42.4 & 27:13:24 & Bipolar & MHO717/718   & 13b & 0.0217 & 8.9 & 0.193 & 172 \\
IRAS~04169+2702     & SE & Class I    & 4:19:58.5 & 27:09:57 & Red     & HH391A/MHO731& 14  & 0.0301 & 6.1 & 0.184 & 112 \\
SE-CO-R1            & SE & --         & 4:19:52.5 & 27:05:30 & Red?    & HH391B/C     &     & 0.0592 & 6.2 & 0.367 & 228 \\
SE-CO-R2            & SE & --         & 4:21:17.1 & 27:02:50 & Red     & MHO720       &     & 0.0021 & 2.6 & 0.005 &   1 \\
J04210795+2702204   & SE & Early TTS  & 4:21:08.0 & 27:02:20 & Bipolar & HH392A       &     & 0.0157 & 5.4 & 0.085 &  46 \\
IRAS~04181+2654A    & SE & Class I    & 4:21:11.5 & 27:01:09 & Red?    &              & 16b & 0.0049 & 2.6 & 0.013 &   3 \\
IRAS~04181+2654B    & SE & Class I    & 4:21:10.4 & 27:01:37 & ?       & HH392B       & 16b?&   --   & --  &  --   & --  \\
SE-CO-flow1$^e$     & SE & --         & 4:20:56.0 & 27:00:00 & Bipolar & HH392C/MHO719&     & 0.0153 & 3.2 & 0.049 &  16 \\
FS~Tau~B            & SE & Early TTS  & 4:22:00.7 & 26:57:32 & Red     & HH157        &     & 0.0209 & 4.9 & 0.102 &  50 \\
CFHT-21$^g$         & SE & TTS        & 4:22:16.8 & 26:54:57 & Bipolar?&              &     &   --   & --  &  --   &  -- \\        
SE-CO-B1$^e$        & SE & --         & 4:22:22.0 & 26:48:00 & Blue    &              &     & 0.0251 & 3.8 & 0.095 &  36 \\

 \hline
 \label{flows}
 \end{tabular}

%\smallskip 
$^a$The most likely driving source of the CO outflow, the area in which it is found in L~1495, 
its YSO type (based on near- and/or mid-IR colours) and its coordinates. \\
If no YSO source is identified the label used for the CO outflow or high-velocity CO peak is 
given.  The coordinates then refer to the approximate midpoint between \\ 
the bipolar lobes, or the peak in the CO emission if the flow is monopolar.  \\
$^b$Indication of whether the CO outflow is bipolar or monopolar (with a red-shifted or 
blue-shifted lobe clearly defined). ``No'' in this column indicates that no CO \\
outflow was detected. \\
$^c$Associated HH objects, MHOs and \hco\ core (from OMK02). \\
$^d$The mass, maximum radial velocity (the difference between the nominal ambient 
velocity and the velocity in the line wings at the 2$\sigma$ noise level); the momentum \\
and the kinetic energy of each outflow lobe. Mean values are quoted if the 
flow is bipolar \\
$^e$CO outflows with no {\em clearly identified} driving source. \\
$^f$Driving source not listed in the YSO catalogue of \citet{luh06}, \citet{ken09}, 
or detected by 2MASS. \\
$^g$CO outflow only a marginal detection; requires confirmation with deeper observations.

\end{tiny}
\end{minipage}
\end{table*}

%%%%%%%%%%%%%%%%%%%%%%%%%%%%%%%%%%%%%%%%%%%%%%%%%%%%%
%%%%%%%%%%%%%%%%%%%%%%%%%%%%%%%%%%%%%%%%%%%%%%%%%%%%%

\subsection{Outflow parameters}

The physical parameters for the molecular outflows described above are
listed in Table~\ref{flows} (we also list HH outflows not detected in
high-velocity CO emission). The mass in the molecular outflow lobes
(the mean of both lobes if the outflow is bipolar) is derived from the
column density integrated across the extent of the high-velocity line
wings, which in turn is calculated from the integrated antenna
temperature, corrected for the main-beam efficiency (we assume a beam
filling factor of unity), and an estimate of the CO 3-2 excitation
temperature.

The excitation temperature, $T_{ex}$, in the ambient gas can be
estimated from the peak temperature in the CO 3-2 line profiles,
provided the gas is optically thick though not self-absorbed
\citep{pin08,buc10}. However, $T_{ex}$ in the ambient gas 
is not necessarily the same as $T_{ex}$ in the entrained outflow gas,
where the kinetic temperature may be higher and the gas is almost
certainly compressed as it is swept up by the underlying jet and
HH/MHO bow shocks \citep[e.g.][]{hat99,dav00a,gia01,lee02,kem09}.
We therefore adopt a somewhat higher value than is implied by (a) the
average CO 3-2 spectra in Fig.~\ref{spec} ($T^*_A \sim 3$~K; $T_{ex}
\sim 11$~K), which include the more diffuse regions away from the
central south-east ridge and star forming cores, or (b) the brightest
spectra observed in L~1495 ($T^*_A \sim 17$~K; $T_{ex} \sim
35$~K)\footnote{In an isothermal slab, assuming local thermodynamic
equilibrium and optically thick emission, the excitation temperature,
$T_{ex}$, is related to $T_{b}$, the line peak main beam brightness
temperature, by: $T_{ex}(3-2) =
16.59/ln[1+16.59/(T_{b}(^{12}CO)+0.036)]$ \citep{pin08}.}.  We instead
use a value of 50~K, consistent with the range in excitation
temperature derived from CO line ratios in other outflows
-- albeit assuming optically thin emission \citep[e.g.][]{dav98,
hat99, dav00a, kem09}.  The mass is in any case relatively insensitive
to temperature, varying by only 40\% over the temperature range
20-100~K \citep{hat07a}.  Assuming an abundance of $X_{\rm CO} =
10^{-4}$ \citep{fre82,wil99}, the column density (in units of
cm$^{-2}$) is then given by $N_{H2} = 2.5 \times 10^{19}(\int T^*_b
dv)$, where the integrated line wings are in units of K~\kms\
\citep{hat07a}.  $\int T^*_b dv$ is measured from the integrated
high-velocity blue-shifted and red-shifted outflow maps used to plot
the contours in Figs.~\ref{co-e}--\ref{co-se4}; the integrated flux
measured in an ellipse that envelopes each outflow lobe is ``sky
subtracted'' using the flux measured in an outer annulus (this
eliminates diffuse red-shifted or blue-shifted gas not associated with
the outflow); the derived column density is then scaled by the HARP
beam size and the factor 1.3$\times m_{\rm H2}$ (where $m_{\rm H2}$ is
the mass of an H$_2$ molecule) to give the mass in each outflow lobe.

We assume that the wings of the CO 3-2 emission are optically thin, so
the values in Table~\ref{flows} will be lower limits: the mass, $M$,
momentum, $P$, and kinetic energy, $E$, are probably underestimated by a
factor of 2-4 \citep{cab90, hat07a}. Furthermore, the radial velocity,
$|v_o-v|$ ($v$ is the velocity in the line wings where the emission
reaches the 2$\sigma$ noise level; $v_o$ is the ambient velocity from
Table~\ref{obs}), used to calculate the momentum and energy, will also
be underestimated, by a factor $1/{\rm sin} i$, where $i$ is the
inclination angle with respect to the plane of the sky. Obviously for
flows close to the plane of the sky this factor can be quite
considerable, although flows with $i \gg$ 60\dg\ will be difficult to
detect in our CO observations; for an inclination angle of 60\dg\ $P$
and $E$ will be underestimated by an additional factor of 2 and 4,
respectively.  Even so, the parameters listed in Table~\ref{flows} are
not unusual for outflows from low-mass young stars.  They are
marginally lower than (within an order of magnitude of) those
typically observed for low-mass YSO outflows in Orion, Ophiuchus and
Perseus
\citep[e.g.][]{kne00,wil03,arc06,bus07,sta07}. 

%%%%%%%%%%%%%%%%%%%%%%%%%%%
%%% Figure %%%%%%%%%%%%%%%%
\begin{figure} 
\hspace*{-0.5cm}
\epsfxsize=9.0cm
%\epsfbox{/export/data/cdavis/jcmt/TAURUS/PAPER-FIGURES/CC-diag.eps}
\epsfbox{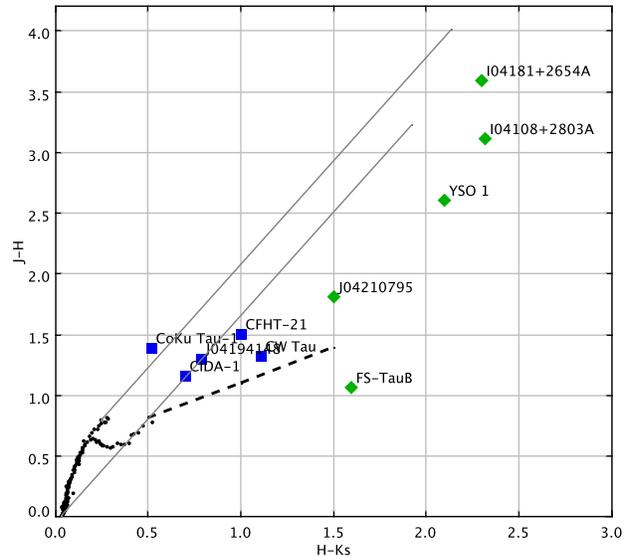}
\caption[] {Near-IR colour-colour diagram for the HH jet and molecular outflow
sources in L~1495, which are marked with blue squares and green
diamonds, respectively.  The black dots lower-left represent the locus
of intrinsic colours for main sequence dwarfs and giants
\citep{koo83}; the dashed line indicates the locus of T Tauri stars
(Meyer, Calvet \& Hillenbrand 1997).  The full parallel lines are
reddening vectors for the main sequence stars up to $A_{\rm v} =
30$~mag \citep{rie85}.}
% we assume a reddening law of the form 
% $R = A_{\rm v}/(E_{\rm B-V}) = 5$ \citep{car89}.  
\label{cc} 
\end{figure}

%%%%%%%%%%%%%%%%%%%%%%%%%%%%%%%%%%%%%%%%%%%%%%%%%%%%%%%%%%%%%%%%%%%%
%%%%%%%%%%%%%%%%%%%%%%%%%%%%%%%%%%%%%%%%%%%%%%%%%%%%%%%%%%%%%%%%%%%%

\section{Discussion}

\subsection{Using outflows to distinguish starless from 
protostellar cores in L~1495}

L~1495 contains the largest concentration of young stars in the Taurus
region.  The majority of these are found in the L~1495 bowl,
particularly the denser eastern half of the bowl \citep[the molecular
gas in the western bowl region is more extensive and generally more
diffuse; ][]{gol08} and in the lower half of the south-east ridge
\citep{luh06,ken09}.  The low number-density of YSOs in the upper,
north-western half of the ridge, where OMK02 nevertheless find a
number of massive, dense cores (9a, 9b, 10a, 10b and 12), suggests
that star formation has yet to take place in this area.  The absence
of HH objects, MHOs, and CO outflows in this region supports
this interpretation.

The molecular outflows in L~1495 are listed in Table~\ref{flows},
along with associated HH objects, MHOs, \hco\ cores and candidate YSO
driving sources.  YSO source classifications are based on near-IR
and/or mid-IR photometry (as described in section 1).  
In Fig.~\ref{cc} we plot the 2MASS colours of the HH jet and outflow
sources in Table~\ref{flows} on a near-IR colour-colour (CC)
diagram.  As expected the majority of the sources lie to the right of
the reddening band (the two parallel lines), although notably the
sources with no CO outflow all lie in the region associated with TTSs,
while the young stars that do drive molecular outflows are all much
redder (in both J-H and H-K$_s$).  Clearly the molecular
outflow-driving sources are more embedded than their HH jet-driving
counterparts.

Of the 22 cores labelled in Fig.~\ref{over} only four -- 13a, 13b, 14
and 16b -- appear to be associated with YSOs (in each case the YSO is
located within the 50\% integrated intensity contour in the \hco\
maps).  This relatively small fraction of cores with YSOs is
consistent with the rest of Taurus; OMK02 find that only 22\% of their
condensations are associated with embedded sources. In L~1495 these
``cores harbouring stars'' certainly seem to be protostellar; at least
three of the four cores (13b, 14 and 16b) are associated with
molecular outflows; the CO flow from 13a is a marginal detection,
though this core is also probably associated with an HH jet and MHO.
As has been noted in other outflow surveys
\citep[e.g.][]{dav08,hat09}, this suggests that CO outflows are useful
tracers of the locations of protostellar cores.  Indeed, outflow
surveys may be used in combination with near- and mid-IR photometric
studies of star forming regions to establish more accurately the
fraction of cores that do harbour accreting protostars.

%%%%%%%%%%%%%%%%%%%%%%%%%%%
%%% Figure %%%%%%%%%%%%%%%%
\begin{figure*}
\centering
\epsfxsize=16.0cm
%\epsfbox{/export/data/cdavis/jcmt/TAURUS/PAPER-FIGURES/Core-plots.eps}
\epsfbox{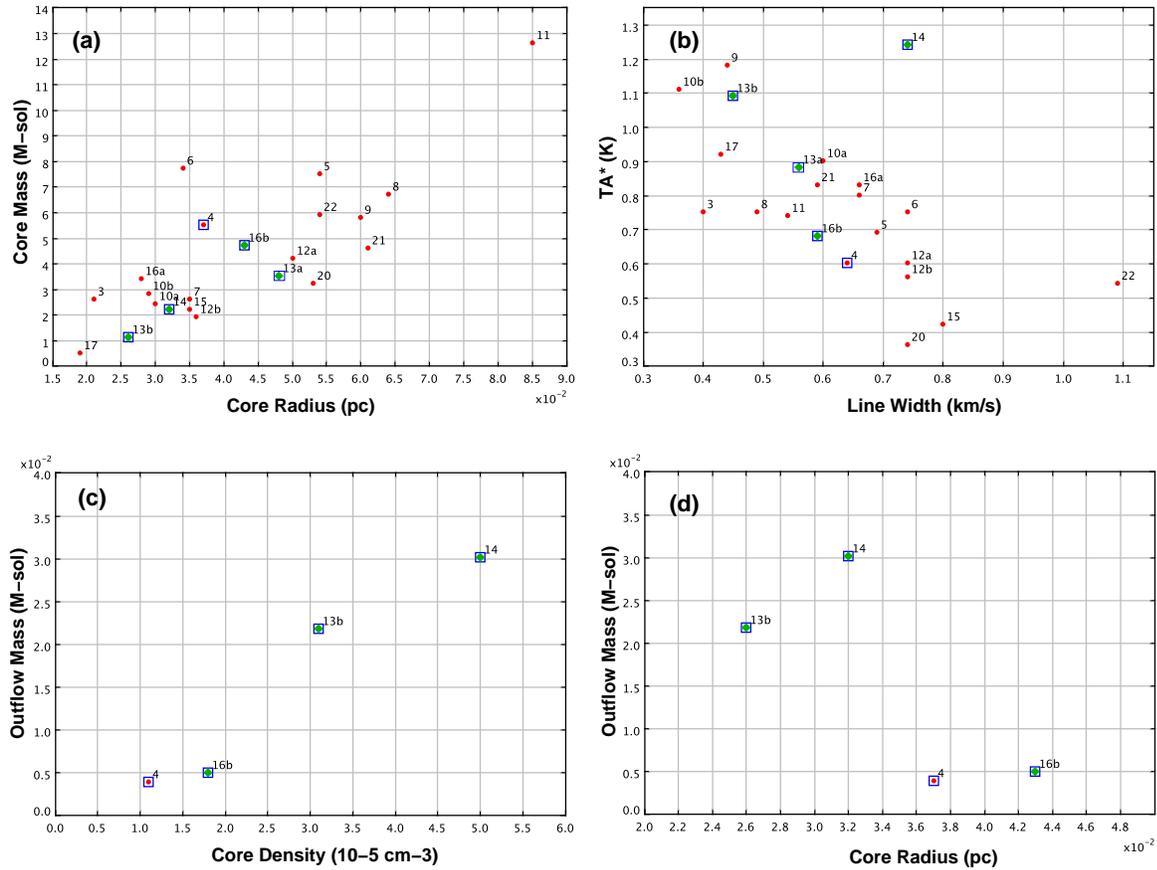}
\caption[] {(a) Plot of \hco\ core mass against core radius; 
(b) plot of \hco\ line width against line intensity measured
towards the peak in each core; (c) and (d) \hco\ core molecular gas
density and core radius plotted against the mass of the associated
outflow.  All core parameters are taken from Table 2 of OMK02. In each
panel, cores are marked with red dots and are labelled according to
OMK02; cores associated with YSOs are marked with green diamonds;
cores associated with molecular outflows or HH jets are marked
with open blue squares (note that we were unable to measure the
molecular flow mass for core 13a). }
\label{cores}
\end{figure*}
%%%%%%%%%%%%%%%%%%%%%%%%%%%%

{\em Spitzer}, when combined with near-IR data from e.g. 2MASS or WFCAM, can
be a powerful tool for searching for Class 0/I protostars (although
saturation can be a problem with {\em Spitzer} data, particularly at
longer wavelengths). This is especially true in nearby, low mass star
forming regions like Taurus, where extinction and crowding are minimal
\citep{eva09}. A few recent studies have revealed the presence of
extremely faint protostellar objects inside a handful of cores that
were previously thought to be starless \citep{cra05,bou06}.  But such
low-mass cores may be below the resolution and sensitivity limits of
the \hco\ survey discussed here, and these objects are unlikely to
drive powerful CO outflows or bright H$_2$ jets. The YSO list, derived
from optical, near-IR and mid-IR photometry and spectroscopy
\citep{ken09}, is therefore expected to be relatively complete for the
core mass range under scrutiny here.  

A number of high-velocity CO features appear to be without driving
sources in Table~\ref{flows}.  In some cases their progenitors may be
nearby; however, in a few (W-CO-flow1, SE-CO-flow1, and
SE-CO-R1/HH~391), the location of the driving source is clearly
defined, yet they are still undetected.  These flows are probably
driven by Class 0 sources which require deeper mid-IR or far-IR/submm
observations \citep{jor07,jor08}.

From our comparison of the \hco\ cores mapped by OMK02 to the YSO
source list and outflows mapped here, at first sight the ratio of
starless cores to protostellar cores in L~1495 seems to be high; as
noted earlier, of the 22 cores located in L~1495, only 4 appear to be
coincident with known YSOs (although all are associated with
outflows). 

However, two cores, 3 and 4, are found in a region that was not
observed with {\em Spitzer}. \citet{luh06} estimate that pre-{\em
Spitzer} studies of the YSO population in Taurus are only complete to
about 80\% . They also point out that {\em Spitzer} is very well
suited to finding disk-bearing stars down to masses of the order of
0.01~\Msol\ through considerable amounts of extinction ($A_{\rm v}
\sim 100$). The observations of OMK02 are sensitive to core masses in the
range 3.5\Msol\ $< M <$ 20.1\Msol .  Even for a very low star
formation efficiency (low protostar-to-core mass ratio) of just a few
percent \citep{war07b}, if these ``starless'' cores did contain
protostars {\em Spitzer} would detect them. Given the small number
statistics here, it is therefore possible that sensitive mid-IR
observations could uncover protostars associated with both cores 3 and
4, in which case six of the 22 cores would be associated with
protostars. (Note also that core 4 may be associated with the CO
outflow W-CO-flow1.)

%% For a low mass
%% Core Mass Function, $dN/dM \sim M^{-2.5}$ (Mott, Andr\'e \& Neri 1998,
%% Testi \& Sargent 1998, OMK02)
%% 
%%At the lower mass limit, the size of a core that forms a low mass
%%YSO...  On the other hand, the number of cores could be also
%%underestimated; the observations of OMK02 are sensitive to core
%%masses in the range 3.5\Msol\ $< M <$ 20.1\Msol . For a low mass
%%Core Mass Function, $dN/dM \sim M^{-2.5}$ (Mott, Andr\'e \& Neri
%%1998, Testi \& Sargent 1998, OMK02)
%%
%%At the lower mass limit, the size of a core that forms a low mass
%%YSO is thought to be of the order of obviously the resolution of the
%%mid-IR observations far exceeds the resolution of the \hco\
%%observations (by an order of magnitude).

Hence, in L~1495 it seems reasonable to assume that we have 5$\pm$1
protostellar cores and 17$\pm$1 starless or prestellar cores; the
ratio of cores that are associated with young stars to those that are
not is in the range 0.2--0.4.  But how many of the starless cores are
prestellar?  OMK02 and \citet{miz94} before them suggest that the
fraction of \hco\ cores in L~1495 that exceed the Jeans mass, are
gravitationally bound, are undergoing some form of collapse, and
therefore are prestellar in nature, is probably high: note that the
\hco\ cores identified by OMK02 are compact (radius $\le 0.1$~pc) and
of high density ($\ge 10^5$~cm$^{-3}$, the critical density for
excitation of the \hco\ line used in their study). They find that
13 of the 22 cores (60\%) in L~1495 have an \hco\ column density mass
that exceeds the virial mass (these 13 include cores 3, 4, 13a, 13b
and 16, i.e. five of the six candidate prestellar cores).  This leads
us to believe that 9 of the 22 are neither prestellar nor
protostellar, so we identify these as being starless.  The ratio of
prestellar to protostellar cores in L~1495 is therefore in the range
$\sim$1.3--2.3, while the ratio of starless to prestellar cores is in
the range $\sim$1--1.3. 

Although based on modest statistics, {\em the above arguments 
indicate that the prestellar phase is about as long lived as the
protostellar phase, and that in L~1495 there are as many prestellar
cores as starless cores. } Similar results have been found in Perseus
\citep{hat07a,jor07,hat08,dav08} and in other low mass star forming regions
\citep{vis02,eno07,eva09}.  

\citet{eva09} find the Class 0 lifetime to be $\sim$10$^5$~yrs and
the Class 0 and Class I lifetimes combined to be about 5 $\times$
10$^5$~yrs. It is worth mentioning that these timescales are
more-or-less consistent with the range in dynamical ages, 10$^4$ --
10$^5$~yrs, measured for molecular outflows (Arce et al. 2007, and
references therein); this is to be expected, given the statistical
evidence that most protostars, particularly those associated with
cores, drive molecular outflows \citep{dav09}.  

\citet{eva09} also find the prestellar core lifetime to be of the
order of 5$\times$10$^5$ yr.  OMK02, from their analysis of the data
discussed here, estimate a very similar timescale ($\sim$
4$\times$10$^5$ yrs) for starless condensations in Taurus.  If we
assume a value of 5$\times$10$^5$ yrs for the protostellar lifetime of
our sources, then we arrive at a prestellar lifetime of
$\sim$9($\pm$3)$\times$10$^5$yrs. This is slightly longer than the
timescale found by Evans et al., although it is still consistent to
within the combined errors.  Furthermore, we find the lifetime of
starless cores that are not gravitationally bound to also be of the
order of 10$^6$~yrs. 

\subsection{The relationship between cores and outflows}

OMK02 estimate various parameters for the \hco\ condensations in
Taurus, including size, density and mass.  We plot a number of these
in Fig.~\ref{cores}: in Fig.~\ref{cores}(a) we chart core mass based
on the integrated intensity in \hco\ emission against core radius (see
OMK02 for details); a plot using the core virial mass, also calculated
by OMK02, looks very similar to this figure.  In Fig.~\ref{cores}(b)
we plot the \hco\ line width against \hco\ peak antenna temperature;
both have been measured by OMK02 at the peak position in each of their
\hco\ cores.  Cores associated with embedded stars and outflows or HH
jets are identified in both figures with green diamonds and open blue
squares, respectively.

The plots in Fig.~\ref{cores}(a) and (b) suggest that the
protostellar, outflow-driving cores are generally less massive, more
compact and more quiescent than many of the observed condensations
(although clearly not all of the compact \hco\ cores are associated
with outflows). At first glance this seems to run against expectation,
since other studies have found that higher mass cores are more likely
to be prestellar \citep[e.g.][]{jor07,hat08}.  However, one should
note that the more massive \hco\ cores are often larger
(Fig.~\ref{cores}(a)) and less dense than their less massive
counterparts; these objects may in fact harbour multiple, unresolved
cores. Alternatively, these massive \hco\ condensations may be younger
than the protostellar cores, or may simply be gravitationally unbound
and therefore not prestellar in nature.

For the \hco\ cores that {\em are} associated with outflows and
protostars, we find no correlation between core mass and outflow mass
or kinetic energy (plots not shown).  If anything, the more massive
cores (4 and 16b) seem to be associated with the least massive flows.
However, this comparison does not take into account the size of each
core; Fig.~\ref{cores}(c) and (d) indicate a possible correlation
between core {\em density} and outflow mass, and an inverse
correlation between core {\em size} and outflow mass, albeit for
a very small number of cores and outflows.  The implication here is
that the build-up of outflow mass increases as the protostellar core
contracts.  This idea clearly requires further investigation, both
observationally and theoretically.  Even so, our results are
consistent with the early analysis of the \hco\ data by
\citet{miz94}, who find that condensations with embedded stars in Taurus 
tend to be smaller in size, more dense, and less extended, than those
without. Similar results were found in the submm continuum
survey of \citet{war94}.

Finally, we have conducted a statistical analysis of the data
plotted in Fig.~\ref{cores} \citep{iso90}.  This suggests that there
is a 93\% chance that the correlation between core mass and core
radius is significant (i.e. there is a 7\% chance that the sample is
random), for cores with stars and for cores with outflows; for the
entire sample there is a much higher chance ($>$99\%) that the
correlation is real, in part because of the much larger number of data
points. Likewise, we find a $>$99\% chance that for all cores the line
intensity is related to the line width (Fig.~\ref{cores}(b)), although
when only cores with stars or cores with outflows are considered the
distribution of points is far less likely to be real ($\sim$65\% and
$<$50\%, respectively). There is a 98.5\% chance that the linear
relationship between outflow mass and core density in
Figs.~\ref{cores}(c) is real, though a reduced likelyhood (85\%) that
the distribution between outflow mass and core mass in
Figs.~\ref{cores}(d) is significant.  Strictly speaking, only the
linear distribution of points in Figs.~\ref{cores}(c) is potentially
real, although clearly both plots would benefit from additional
data. 

%%%%%%%%%%%%%%%%%%%%%%%%%%%%%%%%%%%%%%%%%%%%%%%%%%%%%%%%%%%%%%%%%
%%%%%%%%%%%%%%%%%%%%%%%%%%%%%%%%%%%%%%%%%%%%%%%%%%%%%%%%%%%%%%%%%

\subsection{Star Formation Efficiency in L~1495 based on outflow
statistics}

In their study of outflow activity in Perseus-West, \citet{dav08} note
that the more massive clouds are associated with a greater number of
outflows.  They use cloud masses derived from extinction mapping by
Kirk, Johnstone \& Di Francesco (2006), who note that this is arguably
the best way of tracing the diffuse atomic and molecular gas ($A_v
\sim 3-7$) in large molecular clouds.  Kirk et al. also stress that
only $\sim$5\% of the cloud mass resides in regions with H$_2$ column
densities $> 5\times10^{21}$~cm$^{-3}$ ($A_v > 2.5$): in other
words, only 5\% of the gas mass is locked in the starless and
protostellar cores mapped in \hco\ and in submm dust continuum
emission with, for example, SCUBA or SCUBA-2.  Based on these
measurements, in Perseus-West Davis et al. estimate that there is one
outflow for every 44-88~\Msol\ of ambient material.  This range is not
inconsistent with a canonical value for the star formation efficiency
($SFE = M_{\rm YSO}$/[$M_{\rm YSO} + M_{\rm core}$]) of $\sim$10-15\%
\citep[e.g.][]{jor07}, if the cloud-to-core mass ratio is indeed
$\sim$20, and the protostars have masses of the order of
$\sim$0.5~\Msol.

From their large-scale CO 2-1 and $^{13}$CO 2-1 observations of
Taurus, \citet{gol08} find that the L~1495 bowl and south-east ridge
(regions they label L~1495 and B~213) have masses of the order of
$2.6\times10^3$~\Msol\ and $1.1\times10^3$~\Msol, respectively.  Their
observations are sensitive to H$_2$ column densities in excess of
$\sim 1\times10^{21}$~cm$^{-3}$ ($A_v > 0.5$), so somewhat similar to
the columns of molecular gas probed by the extinction mapping
technique used in Perseus. In L~1495 we list (in Table~\ref{flows})
nine outflows in the bowl and 12 in the L~1495 ridge; we therefore
estimate roughly one outflow per 289\Msol\ in the bowl, and one per
92~\Msol\ of gas in the ridge.  These values are somewhat higher than
in Perseus, especially in the bowl region. This is likely due to the
more evolved state of this region, where the number of outflows
underestimates the number of young stars and therefore the star
formation efficiency. In the L~1495 bowl, a high percentage of young
stars will be weak-line TTSs that are no longer associated with
molecular or even HH flows.  Note also that Taurus in general is
considered to be a less active region of star formation, being
associated with less luminous young stars, narrower molecular line
widths and lower kinetic temperatures than e.g. Perseus or Orion
(Jijina, Myers \& Adams 1999).

%%%%%%%%%%%%%%%%%%%%%%%%%%%%%%%%%%%%%%%%%%%%%%%%%%%%%%%%%%%%%%%%%
%%%%%%%%%%%%%%%%%%%%%%%%%%%%%%%%%%%%%%%%%%%%%%%%%%%%%%%%%%%%%%%%%

\subsection{Outflows, cloud structure and the large-scale B-field in
L~1495}

Taurus has been the subject of many studies of the large-scale cloud
structure and its relationship to the surrounding magnetic field;
\citet{hey88} and \citet{goo92} measure the polarisation of background 
optical and infrared starlight via selective dust absorption,
observations which illustrate the field orientation in the outer, low
column density regions.  \citet{tro96} and \citet{cru00} present
complementary Zeeman effect observations in 18~cm OH emission.
Although these probe higher column densities, they only yield
information on the line-of-sight field component, and are of lower
(arcminute) spatial resolution.  To better trace the B-field that
pervades the dense cores and high-density gas that envelopes these
cores, one requires high spatial resolution observations of polarised
(sub)millimetre continuum emission from magnetically-aligned grains
\citep[e.g.][]{mat01,mat02a,cru04}.

From studies of other low- and intermediate-mass star forming regions
we know that the field orientation, projected onto the plane of the
sky, often snakes through filaments and cores, changing direction on
tenths of parsec scales \citep[e.g.][]{dav00b,mom01,mat02b,hou02}.
Surveys of cores and clumps in star forming clouds also reveal a
general lack of order in the orientations of the long axes of oblate
starless and protostellar cores \citep[e.g.][]{hat05,kir06};
similarly, outflow surveys show that HH and H$_2$ line-emission jets
are usually randomly orientated \citep{sta02,wal05,dav09}. Together,
these data suggest that the magnetic field in the {\em high-density regions}
is poorly coupled to the neutrals, allowing ambipolar diffusion to build
up mass on star forming cores, or that the magnetic energy is
insufficient to overcome the kinetic energy associated with the
dense clumps and cores.  Either way, the orientation of outflows and
their associated protostellar cores do not appear to be strongly
linked to the large-scale B-field.

But could L~1495 be an exception to the rule? The south-east ridge in
L~1495 is notable for being orientated roughly perpendicular to the
surrounding B-field \citep{goo92,hei00,gol08}.  Goodman et al. give a
mean polarisation position angle along the ridge of 27\dg\ from
optical data, and 31\dg\ from infrared polarisation measurements.
\citet{gol08} over-plot polarisation vectors \citep[compiled
by][]{hei00} onto their $^{13}$CO integrated intensity map of Taurus
and find that the field is not only perpendicular to the L~1495 ridge,
but is aligned with ``striations'' in the surrounding low column
density medium.  This suggests that -- in these {\em low density}
regions -- the B-field is well coupled to the molecular gas through
ion-neutral collisions (UV penetration maintaining a degree of
ionisation in the gas).  Collapse along these field lines may then
produce the chain of high density condensations that forms the
south-east ridge in L~1495.  Although the field within the dense ridge
awaits sensitive submm continuum polarisation measurements
(these are planned as part of the Gould Belt survey), we point out
here that many of the outflows found along the south-east ridge are
orientated perpendicular to the ridge, and {\em parallel} with the
field direction in the surrounding low column density gas. This
observation is in general disagreement with the results of Greaves,
Holland \& Ward-Thompson (1997), who in a study of five protostars
with outflows find that if the flow is in the plane of the sky the
field tends to be orientated {\em perpendicular} to the flow axis.
However, Greaves et al. used submm continuum polarisation
measurements and therefore probed the higher column density regions
close to each outflow source.  It will be interesting to see whether
the field direction, measured with the polarimetry facility currently
being developed for SCUBA-2, changes close to the young outflow
sources in L~1495.

%%%%%%%%%%%%%%%%%%%%%%%%%%%%%%%%%%%%%%%%%%%%%%%%%%%%%%%%%%%%%%%%%
%%%%%%%%%%%%%%%%%%%%%%%%%%%%%%%%%%%%%%%%%%%%%%%%%%%%%%%%%%%%%%%%%

\section{Conclusions}

HARP observations in CO 3-2 emission are shown to be ideal for tracing
outflow activity in nearby star forming regions.  The observations
discussed here reveal as many as 16 molecular outflows in L~1495; most
are associated with HH objects and/or molecular hydrogen line-emission
features (MHOs).  Candidate outflow driving sources (protostars or
TTSs) are identified for eight CO flows, although only four of these
flow progenitors appear to be associated with \hco\ cores.  Even so, we
note that the CO outflow-driving sources have redder near-IR colours
than their HH jet-driving counterparts.  We also find a possible
correlation between outflow mass and the associated core density and
size, the more massive flows being associated with the denser, more
compact cores.

OMK02 find that 13 of the 22 \hco\ cores in L~1495 have a
column-density mass that exceeds the virial mass.  Of these 13 cores,
five seem to be associated with protostars (and four of these drive
outflows).  We estimate that the ratio of prestellar to protostellar
cores in L~1495 is approximately in the range 1.3--2.3; the ratio of
starless to prestellar cores is estimated to be $\sim$1.
 
Overall, we find that the bowl is more evolved than the south-east
ridge in L~1495; in the bowl there is a paucity of molecular flows
though a larger fraction of TTSs.  The fact that the ridge is
spatially more compact (long and narrow) in comparison to the bowl
supports this interpretation.  The star formation efficiency in L~1495
(particularly the south-east ridge), estimated from the ambient cloud
density, a canonical value for the cloud-to-core mass ratio, and the
observed number of outflows, is consistent with other low-mass star
forming regions.

In comparison to Orion and Perseus there is a modest number of
outflows and protostars in L~1495.  However, the region, especially
the south-east ridge, is relatively simple (in comparison to, say,
NGC~1333), with little chance for source confusion and outflows
overlapping other outflows or unrelated molecular cores. The
statistics, though modest, are therefore likely to be more robust than
in other regions.  

%%%%%%
%%%%%%

\section*{Acknowledgements}

We thank the anonymous referee for his/her comments, which improved
the overall quality of this paper.  The James Clerk Maxwell Telescope
is operated by the Joint Astronomy Centre (JAC) on behalf of the
Science and Technology Facilities Council (STFC) of the United
Kingdom, the Netherlands Organisation for Scientific Research, and the
National Research Council of Canada.  UKIRT is operated by the JAC on
behalf of the STFC.  We acknowledge the Cambridge Astronomical Survey
Unit (CASU) and the WFCAM Science Archive (WSA) for the processing and
distribution of the near-IR data presented in this paper. This
research used the facilities of the Canadian Astronomy Data Centre
operated by the National Research Council of Canada with the support
of the Canadian Space Agency.

%%%%%%%%%%%%%%%%%%%%%%%%%%%%%%%%%%%%%%%%%%%%%%%%%%%%%%%%%%%%%%%%%
%%%%%%%%%%%%%%%%%%%%%%%%%%%%%%%%%%%%%%%%%%%%%%%%%%%%%%%%%%%%%%%%%

\end{document}